\documentclass[a4paper,11pt]{article}
\pdfoutput=1
\usepackage{graphicx,rotating,hyperref,slashed,amsmath,xcolor,amssymb,amsfonts,colortbl,cite} 
\makeatletter  
\hypersetup{colorlinks,bookmarksopen,bookmarksnumbered,
linkcolor=blus,pdfstartview=FitH,urlcolor=rossos,citecolor=verde}
\allowdisplaybreaks	 
\numberwithin{equation}{section} 
\hypersetup{%
    ,urlcolor=black
    ,citecolor=black
    ,linkcolor=black 
    }
\usepackage{mciteplus}

\AtBeginDocument{% optionally set colors to your liking
  \hypersetup{
    urlcolor=blue,
    citecolor=blue,
    linkcolor=black,
  }%
}

\def\beq{\begin{equation}}
\def\eeq{\end{equation}}

\newcommand\ees{\end{eqnarray}}
\newcommand\bees{\begin{eqnarray}}

\def\bea{\begin{eqnarray}}
\def\eea{\end{eqnarray}}

\def\d{{\rm d}}

\def\d{{\rm d}}

\def\0{{\boldsymbol 0}}

\def\lsim{\mathrel{\rlap{\lower3pt\hbox{\hskip0pt$\sim$}}
   \raise1pt\hbox{$<$}}}         %less than or approx. symbol
\def\gsim{\mathrel{\rlap{\lower4pt\hbox{\hskip1pt$\sim$}}
   \raise1pt\hbox{$>$}}}         %greater than or approx. symbol

 \newcommand{\sfootnote}[1]{}
\definecolor{bluc}{cmyk}{1,1,0,0.1}
\definecolor{rossoCP3}{cmyk}{0,.88,.77,.40}
\definecolor{rosso}{cmyk}{0,1,1,0.4}
\definecolor{rossos}{cmyk}{0,1,1,0.55}
\definecolor{rossoc}{cmyk}{0,1,1,0.2}
\definecolor{verdes}{cmyk}{0.92,0,0.59,0.4}

\hypersetup{colorlinks, bookmarksopen, bookmarksnumbered,
citecolor=verdes, linkcolor=bluc, pdfstartview=FitH, urlcolor=rossos}

\usepackage{multicol}
\usepackage{color}
\definecolor{rosso}{cmyk}{0,1,1,0.4}
\definecolor{rossos}{cmyk}{0,1,1,0.55}
\definecolor{rossoc}{cmyk}{0,1,1,0.2}
\definecolor{blu}{cmyk}{1,1,0,0.3}
\definecolor{blus}{cmyk}{1,1,0,0.6}
\definecolor{bluc}{cmyk}{1,1,0,0.1}
\definecolor{verde}{cmyk}{0.92,0,0.59,0.25}
\definecolor{verdec}{cmyk}{0.92,0,0.59,0.15}
\definecolor{verdes}{cmyk}{0.92,0,0.59,0.4}

\skip\@mpfootins = \skip\footins
\def\circa#1{\,\raise.3ex\hbox{$#1$\kern-.75em\lower1ex\hbox{$\sim$}}\,}

\newcommand{\be}{\begin{equation}}
\newcommand{\ee}{\end{equation}}
\newfam\rsfsfam
\def\mathscr#1{{\fam\rsfsfam\relax#1}}

\def\circa#1{\,\raise.3ex\hbox{$#1$\kern-.75em\lower1ex\hbox{$\sim$}}\,}
\makeatletter

\def\hhref#1{\href{http://arxiv.org/abs/#1}{arXiv:#1}} % in bibliography

\newcommand{\doi}[1]{\href{http://dx.doi.org/#1}{[doi]}}

\def\hhref#1{\href{http://arxiv.org/abs/#1}{arXiv:#1}} 
 
\def\art{\@ifnextchar[{\eart}{\oart}}
\def\eart[#1]#2#3#4#5#6{{\rm #2}, {\em #3 \bf #4} {\rm (#6) #5} ({\em #1})}

\def\article{\@ifnextchar[{\earticle}{\oarticle}}
\def\oarticle#1#2#3#4#5#6{{\rm #1}, {\em ``#6''}, {\rm #2 #3 (#5) #4}}
\def\earticle[#1]#2#3#4#5#6#7{{\rm #2}, {\em ``#7''}, {\rm #3 #4 (#6) #5}  [\hhref{#1}]}
\def\hepart[#1]#2{{\rm #2, \em#1}}
\def\heparticle[#1]#2#3{#2, {\em ``#3''} [\hhref{#1}]}

\newcounter{alphaequation}[equation]
\def\thealphaequation{\theequation\hbox to
0.6em{\hfil\alph{alphaequation}\hfil}}
\def\eqnsystem#1{
\def\@eqnnum{{\rm (\thealphaequation)}}
\def\@@eqncr{\let\@tempa\relax \ifcase\@eqcnt \def\@tempa{& & &} \or
  \def\@tempa{& &}\or \def\@tempa{&}\fi\@tempa
  \if@eqnsw\@eqnnum\refstepcounter{alphaequation}\fi
\global\@eqnswtrue\global\@eqcnt=0\cr}
\refstepcounter{equation} \let\@currentlabel\theequation \def\@tempb{#1}
\ifx\@tempb\empty\else\label{#1}\fi
\refstepcounter{alphaequation}
\let\@currentlabel\thealphaequation
\global\@eqnswtrue\global\@eqcnt=0 \tabskip\@centering\let\\=\@eqncr
$$\halign to \displaywidth\bgroup \@eqnsel\hskip\@centering
$\displaystyle\tabskip\z@{##}$&\global\@eqcnt\@ne
\hskip2\arraycolsep\hfil${##}$\hfil& \global\@eqcnt\tw@\hskip2\arraycolsep
$\displaystyle\tabskip\z@{##}$\hfil
\tabskip\@centering&\llap{##}\tabskip\z@\cr}
\def\endeqnsystem{\@@eqncr\egroup$$\global\@ignoretrue} \makeatother

\definecolor{fiorentina}{rgb}{.5,0,.5}

\begin{document}
%\preprint{ET-0465A-21}

%ET-0465A-21
\setcounter{page}{1} \baselineskip=15.5pt \thispagestyle{empty}

\vspace{0.8cm}
\begin{center}

{\fontsize{19}{28}\selectfont  \sffamily \bfseries {
Non-Gaussianities and the large $|\eta|$ approach to  inflation}}

\end{center}

\vspace{0.2cm}

\begin{center}
{\fontsize{13}{30}Gianmassimo Tasinato$^{1,2}$ 
} 
\end{center}

\begin{center}
{\small
\vskip 8pt
\textsl{$^{1}$ Physics Department, Swansea University, SA28PP, United Kingdom}\\
\textsl{$^{2}$ Dipartimento di Fisica e Astronomia, Universit\`a di Bologna, and }\\
\textsl{INFN, Sezione di Bologna, I.S. FLAG, viale B. Pichat 6/2, 40127 Bologna, Italy}\\
\vskip0.1cm
\textsl{\texttt{email}: g.tasinato2208 at gmail.com }\\
\vskip 7pt
}
\end{center}
\smallskip
\begin{abstract}
\noindent
The physics of primordial black holes  can be affected
by the non-Gaussian statistics of the density fluctuations
that generate them. Therefore,   it is important to have  good theoretical
control of the  higher-order correlation functions for primordial curvature perturbations. 
By working at leading order in a $1/|\eta|$ expansion,
we analytically determine the bispectrum
of curvature fluctuations for single field   inflationary scenarios producing primordial black holes. 
The bispectrum has a rich scale and shape dependence, and its features
depend on the dynamics of the would-be decaying  mode. 
We apply our analytical results to study  
 gravitational waves induced at second order by enhanced
curvature fluctuations. 
 Their  statistical properties are derived in terms of convolution integrals 
 over wide  momentum ranges, and they are sensitive on the scale and shape dependence
  of the curvature bispectrum we analytically computed.
\end{abstract}

%%%%%%%%%%%%%%%%%%%%%%%%%%%%%%%%%%%%%%%%%%%%%%%%%%%%%%%%%%%
\section{Introduction and Conclusions}
\label{sec_intro}
%%%%%%%%%%%%%%%%%%%%%%%%%%%%%%%%%%%%%%%%%%%%%%%%%%%%%%%%%%%%%%%%%%%%%%%%%%%%%%%%%%%%%%%

The lack of direct detection   of particle dark matter motivates the study 
  of primordial black holes (PBH) \cite{Hawking:1971ei,Carr:1974nx,Ivanov:1994pa,Garcia-Bellido:1996mdl} as dark matter candidate. Primordial
  black holes are formed by the collapse of  density fluctuations produced during cosmic inflation. See e.g. \cite{Khlopov:2008qy,Garcia-Bellido:2017fdg,Sasaki:2018dmp,Carr:2020xqk,Green:2020jor,Escriva:2022duf,Ozsoy:2023ryl} for reviews. In order for producing PBH, the size of the inflationary curvature fluctuation spectrum  should increase by several  orders of magnitude from large towards small scales. 
  This condition can be achieved by  models of inflation  violating   the standard slowroll conditions, at least during a short
 epoch within the inflationary process  (see e.g. \cite{Baumann:2009ds} for an introduction to standard slowroll inflation). In the simplest PBH-forming scenarios, while the parameter $\epsilon$ remains small, the 
  absolute value of the parameter  $\eta$  become well  larger than one during a short range
  $\Delta N_{\rm NSR}$ of inflationary e-folds.
  Typically,
  since
  we can not rely on a perturbative slowroll  expansion,
   the analysis of such non slowroll models requires the use of numerical techniques   --  
  although interesting specific scenarios exist, which are amenable of analytic investigation: see e.g. \cite{Starobinsky:1992ts,Wands:1998yp,Leach:2001zf,Ozsoy:2019lyy,Carrilho:2019oqg,Tasinato:2020vdk,Franciolini:2022pav,Karam:2022nym,Pi:2022zxs,Domenech:2023dxx}. 
  
  The work  \cite{Tasinato:2023ukp} proposes a new perturbative framework based on an expansion in inverse powers of a small parameter $1/|\eta|$. Let us   assume that the quantity $|\eta|$ is very large. 
   At the same time, 
  the 
  duration   of non slowroll phase $\Delta N_{\rm NSR}$ is infinitesimally small, 
  while the value for the product $|\eta|\, \Delta N_{\rm NSR}$ is kept finite.  
Cosmological observables based on correlators of curvature perturbations can be then calculated   analytically, leading to   formulas
  organized  in a $1/|\eta|$ expansion. 
   There is hope
 that the results of these calculations   share general features with the statistics
 of curvature perturbations   computed within more general (and realistic) families
 of  PBH-forming scenarios, characterised by finite values of $|\eta|$. There
 are interesting  analogies with other approaches developed within  quantum field theory, as  't Hooft  $1/N$
 expansion \cite{tHooft:1973alw}. 
  In such a framework,
  the large $N$ limit of $SU(N)$ gauge theories is able to 
 shed light on the properties of quantum chromodynamics, characterized by  $N=3$ (see e.g. \cite{Coleman:1985rnk} for a pedagogical introduction to the subject). 
 
 \smallskip
 
 The scope of this work is to analyze more systematically  features of non-Gaussian
 correlators using the large $|\eta|$ approach to single field inflation, going beyond the investigations
  started  in \cite{Tasinato:2023ukp}.  
Primordial  Non-Gaussianities %{\bf \color{red} here extra refs}
    can  
   be generated in scenarios producing PBH in inflation, see e.g. \cite{Garcia-Bellido:2016dkw,Garcia-Bellido:2017aan,Ezquiaga:2019ftu,Pi:2021dft,Cai:2021zsp,Pi:2022ysn}. 
   They 
    play an important role in the physics of primordial black holes, 
    since the formation of compact objects depends on the statistics
    of curvature fluctuations and their probability distribution function, 
  see e.g. \cite{Saito:2008em,Byrnes:2012yx,Young:2013oia,Bugaev:2013vba,Young:2015kda,Franciolini:2018vbk,Passaglia:2018ixg,Atal:2018neu,Biagetti:2018pjj,Atal:2019cdz,Kehagias:2019eil,DeLuca:2019qsy,Young:2019yug,Germani:2019zez,Kawasaki:2019mbl,Yoo:2019pma,Figueroa:2020jkf,Taoso:2021uvl,Hooshangi:2021ubn,Young:2022phe,Ferrante:2022mui,Gow:2022jfb,Tomberg:2023kli,Li:2023xtl,Franciolini:2023pbf}. 
It is then important to have a good analytical control of the properties
  of the primordial bispectrum in PBH models.  
  After a brief review of the motivations
 and tools for  carrying on a $1/|\eta|$ expansion -- see section \ref{sec_ps}, based on \cite{Tasinato:2023ukp} -- we focus in section \ref{sec_bis} in
 computing the bispectrum. We 
  discuss the properties of the three point functions of curvature perturbations.  Interestingly, our approach allows us to  determine  the leading contributions of the third order interactions
  Hamiltonian,
  % that give contributions in the   large $|\eta|$ limit, It    unambiguously fixes all the coefficients needed to perform  the computation, 
   and leads us to analytically compute
  the bispectrum at leading order in a $1/|\eta|$ expansion. The result depends on a {\it single} free
  parameter, which controls the enhancement of the spectrum from large towards small scales.  
  We nevertheless find a very rich scale and shape dependence, which we analyze in  detail. Depending on the scale, the bispectrum can be enhanced on an equilateral shape, or on more elongated shapes. In the squeezed limit, the
  bispectrum satisfies  Maldacena consistency relation, while in a not-so-squeezed regime it can enhance the effects of the non slowroll era. For the equilateral shape, the    scale dependence of the bispectrum unexpectedly   resembles the profile found in section \ref{sec_ps} for the power spectrum. We interpret such behaviours as due to the dynamics of the would-be decaying
  mode, which affects in similar ways both the two and three point functions for curvature fluctuations. 
 
 \smallskip
 
 Armed with these analytical results, we   study in section \ref{sec_gw} specific  observables that are sensitive 
 to the shape and scale dependence of the bispectrum. 
 We focus on correlation
 functions  involving scalar induced gravitational waves. Gravitational wave tensor modes
 can be sourced at second order in perturbations by scalar fluctuations
 amplified at small scales.  The amplitude and properties
 of the induced tensor fluctuations can be computed in terms of convolution integrals, 
 involving the statistics of the scalar modes sourcing them. Such convolution
 integrals are sensitive to the properties of curvature fluctuations at all scales, not only 
 on scales 
 nearby
 the peak of the curvature power spectrum.
 See e.g. \cite{Matarrese:1992rp,Matarrese:1993zf,Ananda:2006af,Baumann:2007zm,Saito:2008jc,Saito:2009jt,Espinosa:2018eve,Domenech:2019quo} for some of the original papers, \cite{DeLuca:2019ufz,Inomata:2019yww} for discussions on
 gauge issues on this framework,
 and \cite{Domenech:2021ztg} for a comprehensive review (and
 references therein). The primordial non-Gaussianity of curvature fluctuations  can affect the correlation  functions
 of induced gravitational waves  \cite{Cai:2018dig,Unal:2018yaa,Yuan:2020iwf,Atal:2021jyo,Adshead:2021hnm,Chang:2022nzu,Garcia-Saenz:2022tzu,Li:2023qua}.  Most studies  in the literature assume a local form for primordial non-Gaussianity. It is important
 to extend the analysis to more realistic cases of full  bispectra computed in specific scenarios, as  the ones we determine
 in section \ref{sec_bis}. 
   In fact,  a good knowledge of the shape and scale dependence of the scalar bispectrum through all scales is  essential when evaluating convolution integrals.
   We consider a non-Gaussian observable correlating two (induced) tensor and one scalar mode.  Such observable, in the squeezed limit, plays a role for the multimessenger cosmology proposal pushed forward in  \cite{Adshead:2020bji}. We find  an analytical expression for this quantity.
   % which  makes
% manifest 
 %the role of the shape and scale  dependence of the scalar bispectrum into the computation. 
 The resulting induced non-Gaussianity has a strong scale dependence in the squeezed configuration, with features
 enhanced around the position of the peak of the induced tensor power spectrum. The overall magnitude of the result is  small, but our findings set the stage for more complete analytical studies 
 of the effects of non-Gaussianities of general form  in the computations of  the statistics of induced gravitational
 waves.

  \smallskip
  Our analysis can be extended along several directions. The  large $|\eta|$ formalism can be applied to
  compute the  four point function for scalar curvature fluctuations, which can  hopefully be reconstructed
  analytically and unambiguously  starting from the fourth order interaction Hamiltonian for
  perturbations in single field inflation. The result would be important to compute the trispectrum contributions to the induced gravitational wave spectrum (see e.g. \cite{Adshead:2021hnm,Garcia-Saenz:2022tzu,Maity:2023qzw}), in a realistic set up which include the complete shape and scale dependence
  of the quantities involved. Moreover, the  large $|\eta|$ approach can be helpful for the ongoing
  debate of one-loop corrections in PBH-forming scenarios \cite{Kristiano:2022maq,Kristiano:2023scm}, which is so far unresolved, see e.g.  \cite{Riotto:2023hoz,Choudhury:2023vuj,Choudhury:2023jlt,Riotto:2023gpm,Choudhury:2023rks,Firouzjahi:2023aum,Motohashi:2023syh,Choudhury:2023hvf,Firouzjahi:2023ahg,Firouzjahi:2023btw,Franciolini:2023lgy,Fumagalli:2023loc,Tada:2023rgp,Firouzjahi:2023bkt}. In \cite{Tasinato:2023ukp} we shown that one loop corrections can be set under control at large scales in the large $|\eta|$ limit, without including the boundary terms later discussed in \cite{Fumagalli:2023loc}
  in the context of  loop corrections (but see also \cite{Firouzjahi:2023bkt}). Once the debate on the correct
  way to carry on the computations will be fully clarified, a  consistent analytical investigation in the large $|\eta|$
  limit  will  certainly be instructive. 
  
  %%%%%%%%%%%%%%%%%%%%%%%%%%%%%%%%%%%%%%%%%%%%%%%%%%%%%%%%%%
\section{Mode functions and  curvature power spectrum}
\label{sec_ps}
%%%%%%%%%%%%%%%%%%%%%%%%%%%%%%%%%%%%%%%%%%%%%%%%%%%%%%%%%%

Consider an  inflationary scenario that undergoes a brief, but drastic phase of slowroll violation. The cosmic evolution  corresponds to   quasi de Sitter expansion, with a spacetime characterized by a
 conformally flat FLRW metric
\be
d s^2\,=\,a^2(\tau)\,\left(-d \tau^2+d \vec x^2 \right)\,.
\ee
The scale factor is $a(\tau)\simeq -1/(H_0 \tau)$ for negative conformal time, and nearly constant Hubble parameter $H_0$.  Inflation occurs
at negative conformal time, and ends at $\tau=0$.
 The first slowroll parameter $\epsilon\,=\,-d \ln H/(d \ln a)$
 remains small. But
 it decreases rapidly during a short phase of non slowroll evolution, during which the second
 slowroll parameter $\eta={d \ln \epsilon}/{d \ln a} $ is negative and large in absolute value.
 Scenarios of  ultra slowroll inflation \cite{Kinney:2005vj,Martin:2012pe,Dimopoulos:2017ged} where $\eta=-6$ represent typical  situations where the $|\eta|$ parameter is well larger than one for a brief period. They are widely used when discussing the physics of primordial
 black holes (see e.g. \cite{Ozsoy:2023ryl} for a review on model building). Other scenarios 
 are constant roll models \cite{Motohashi:2014ppa,Inoue:2001zt,Tzirakis:2007bf}, and include specific constructions with 
 with arbitrarily large values of   $|\eta|$, see e.g. \cite{Inomata:2021tpx}. In this work, following \cite{Tasinato:2023ukp}, we formally consider a case in which  
 $|\eta|$ is  large, and at the same time the duration of non slowroll phase is brief. 
 We indicate  with $\tau_1$ and $\tau_2$ respectively
 the (negative) conformal times when the non slowroll phase starts and ends. 
 We define the combination
 \be
 \label{defDta}
 \Delta \tau\,=\,-\frac{\tau_2-\tau_1}{\tau_1}\,,
 \ee
 and consider
 the large $|\eta|$ limit:
 \be
 \label{defle}
 %{\text{large $|\eta|$ limit:\,\,}} \hskip0.5cm
  |\eta|\to\infty \hskip0.5cm  {\rm and} \hskip0.5cm  \Delta \tau\to0\,,  \hskip0.5cm  {\rm while} \hskip0.3cm |\eta|\,\Delta \tau\,=\,\frac{\Pi_0}{2}   \hskip0.3cm  {\rm remains\,\, finite.}
 \ee

 The quantity $\eta$ is computed at the onset of the non slowroll phase.
 This limit 
  is well defined, and leads to  meaningful expressions for physical quantities.
  The latter
  depend  on the single parameter $\Pi_0$, which  encapsulates
 all the effects of the brief non slowroll era. The
 condition \eqref{defle} suggests a consistent
 perturbative expansion in inverse powers of $|\eta|$: whenever
 we find a $\Delta \tau$ in our formulas, we substitute it with $\Pi_0/(2 |\eta|)$
 following the prescription of  eq \eqref{defle}, to then perform an expansion in the small $1/|\eta|$ parameter.
 In analogy with 't Hooft large $N$ expansion
 \cite{tHooft:1973alw}, we dub the limit of eq \eqref{defle} 
 {\it
 inflationary large $|\eta|$ expansion} (see \cite{Tasinato:2023ukp} for details). 
 
 \smallskip 
The physical quantities we will be interested in are correlators of primordial
 fluctuations. We define a dimensionless quantity 
\be
\label{defkap}
- k \tau_1\,=\,\kappa\,
\ee
combining  the Fourier scale $k$ with the conformal time $\tau_1$
around which the non slowroll era occurs.
The value
 $\kappa \sim {\cal O}(1)$ is the typical scale of modes leaving the horizon during  the non slowroll era. 
The evolution of the scalar curvature perturbation $\zeta_{\kappa}(\tau)$
 in Fourier space  obeys the Mukhanov-Sasaki equation (see e.g. \cite{Baumann:2009ds} for a textbook discussion).
 
 %\ee
 As anticipated above, we  assume that  the evolution satisfies slowroll conditions
 up to a brief phase $\tau_1\le\tau\le \tau_2$. In the limit of small $\Delta \tau$, the expression for the mode functions can be found analytically \cite{Tasinato:2020vdk}. See Appendix \ref{appA} for
 a  review. Starting from formulas \eqref{finm2}-\eqref{expb2}, and taking the limit \eqref{defle}, we
 find  the solution 
 for the  mode function satisfying the Mukhanov-Sasaki equation
 at times $ \tau_2\le \tau\le0$ after the non slowroll phase ends: 
  \bea
\zeta_{ \kappa}(\tau)&=&-\frac{i\,H_0\,(-\tau_1)^{3/2}}{2\,\sqrt{\epsilon_1}\,\kappa^{3/2}}\,e^{i \kappa \tau/\tau_1} \left(1-\frac{i \kappa\,\tau}{\tau_1}\right)+\nonumber\\
&-&\frac{i\,H_0\,(-\tau_1)^{3/2}\,\Pi_0}{2\,\sqrt{\epsilon_1}\,\kappa^{5/2}} \,e^{2 i \kappa-\frac{i \kappa \tau}{\tau_1}}\,\left[\left(1-{i \kappa}\right)
 \left(1+\frac{i \kappa\,\tau}{\tau_1}\right) +
e^{-2 i \kappa+ \frac{2 i \kappa \tau}{\tau_1}} \left(1+{i \kappa}\right)
 \left(1-\frac{i \kappa\,\tau}{\tau_1}\right) \right]\,,
 \nonumber
 \\
  \label{solzm}
\eea
where $\epsilon_1$ is the constant, small  parameter
defined in the first phase of inflationary slowroll evolution $\tau\le\tau_1$. 
All the effects in the mode functions  of a non slowroll, large $|\eta|$ era 
are contained in the parameter $\Pi_0$ defined in eq \eqref{defle}, which multiplies
the second line of eq \eqref{solzm}.

\begin{figure}
\begin{center}
    \includegraphics[width = 0.5 \textwidth]{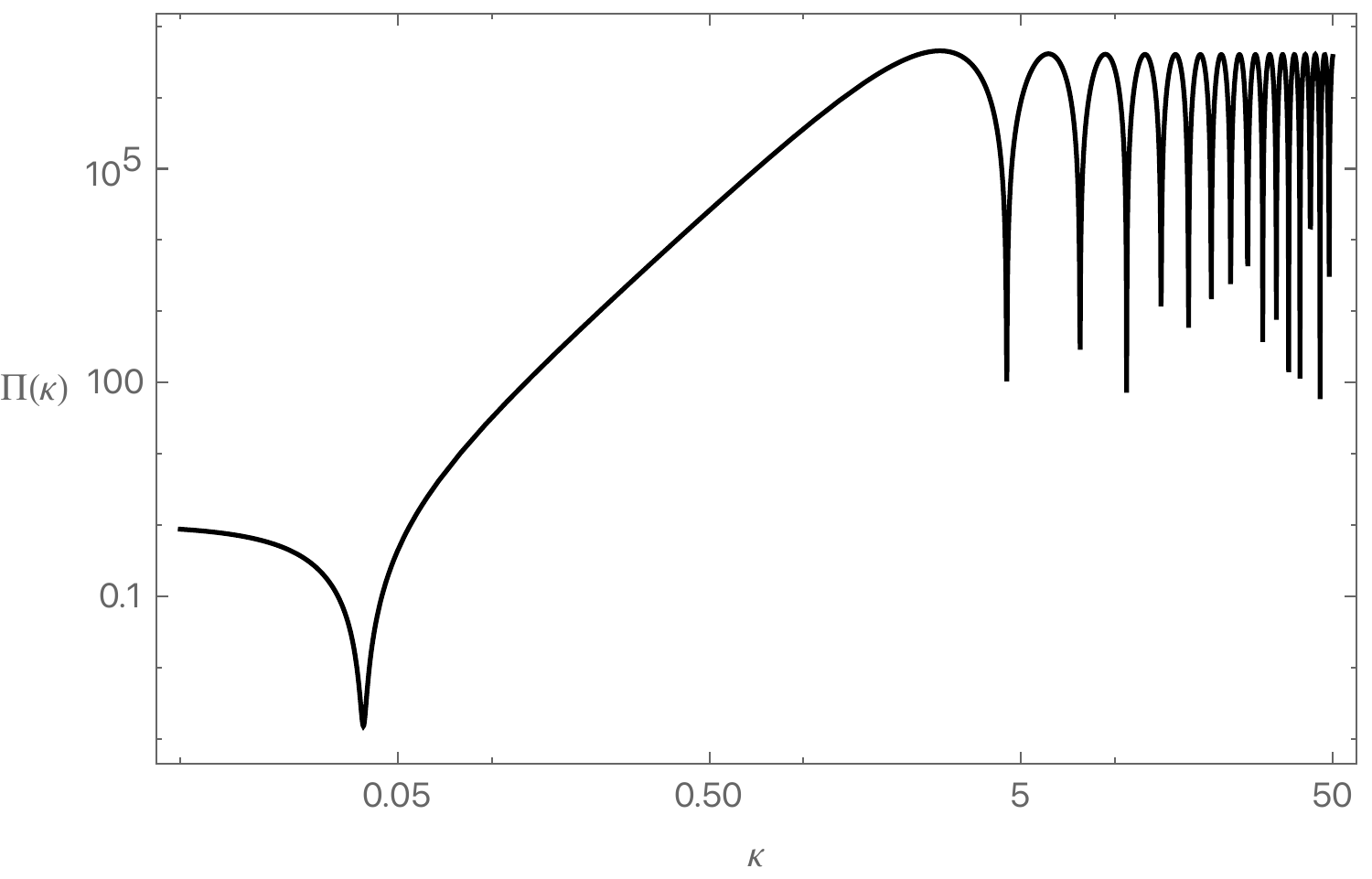}
\caption{\footnotesize The quantity $\Pi(\kappa)$ introduced in eq \eqref{defPa}, evaluated for $\Pi_0=10^3$.}
  \label{lab_fig1}
    \end{center}
\end{figure}

Starting from eq \eqref{solzm}, we review for the rest
of this section some of the results of \cite{Tasinato:2023ukp} on the  
power spectrum of  the curvature perturbation,  evaluated
at the end of inflation $\tau=0$.  In the next section, then, we 
 analyse the bispectrum.
The power spectrum at the end of inflation is defined in terms of two point correlators of curvature perturbation 
\be
\label{gedeP}
\langle \zeta_{ \kappa}(\tau=0) \zeta_{ \rho}(\tau=0)  \rangle\,=\, \delta(\vec \kappa+\vec \rho)
\,P_\kappa\,.
\ee
At very large scales, $\kappa\ll1$, the power spectrum acquires the usual
scale invariant limit
\be
P_{\kappa\ll1}(\tau)\,=\, \frac{{\cal P}_0}{\kappa^3}\,, \hskip0.5cm
{\rm with}\hskip0.5cm {\cal P}_0\,=\,\frac{H_0^2\,(-\tau_1)^3}{4 \epsilon_1}\,.
\label{defsci}
\ee
However, at smaller scales, the structure of the mode function \eqref{solzm}
leads to a rich scale dependence for the two point function.
 We define the dimensionless
spectrum evaluated at the end of inflation:
\be
\label{defPa}
\Pi(\kappa)\,=\,\frac{P_{\kappa}(\tau=0)}{P_{\kappa\ll1}(\tau=0)}\,,
\ee 
as a ratio between values of the spectrum at scale $\kappa$, versus
the the spectrum at $\kappa=0$. 
 Eq \eqref{solzm} leads to
\be
\label{resPIa}
{ \Pi}(\kappa)\,=\,
1+4 \Pi_0 \,\kappa\, j_1(\kappa) \left( 
\kappa j_1(\kappa)-j_0(\kappa)
\right)+4\,\Pi_0^4\,\kappa^2 j_0^2 (\kappa)\,,
\ee
where
\bea
j_0(\kappa)&=&\frac{\sin \kappa}{\kappa}
\hskip0.7cm;\hskip0.7cm
j_1(\kappa)\,=\,\frac{\sin \kappa}{\kappa^2}-\frac{\cos \kappa}{\kappa}
\label{defbes}
\eea
are the spherical Bessel functions. The profile for the function ${\Pi}(\kappa)$ is represented
in Fig \ref{lab_fig1}. As anticipated, in the limit of infinitely large $|\eta|$, the $1/|\eta|$
corrections vanish and  the resulting spectrum depends only on one
parameter, $\Pi_0$. Corrections of order $1/|\eta|$
and higher can be computed starting with formulas of Appendix \ref{appA}. They improve the small scale behaviour reducing the amplitude of oscillations for $\kappa>1$ in Fig \ref{lab_fig1}, but they do not affect the large scale profile $\kappa\le1$.  It is clear that the amplitude of the spectrum
increases considerably from large towards small scales, reaching
its maximal values starting around $\kappa \sim {\cal O}(1)$. In fact, the parameter $\Pi_0$ 
in eq \eqref{defle}
has a transparent physical
interpretation, since it controls the amplification  of the spectrum between large and small scales (see \cite{Tasinato:2023ukp} and Appendix \ref{appA}):
\be
\frac{\lim_{\kappa \to \infty} P_{\kappa}(\tau=0)}{\lim_{\kappa \to 0} 
P_{\kappa}(\tau=0)}
\,=\,
\left(1+\Pi_0 \right)^2\,.
\label{infre}
\ee
If we wish a large value of $ \Pi(\kappa)$
at small scales, as required in models leading to primordial
black hole formation, then the value of the parameter $\Pi_0$ should be
large. We assume this condition in what follows. 

What is causing the rich scale dependence of the curvature
spectrum, as represented in Fig \ref{lab_fig1}? The responsible is  the would-be
decaying mode, that does not actually decay at  superhorizon scales
  during
the phase of non slowroll evolution. It is the disruptive interference
between decaying and non decaying mode that produces  the dip in the 
spectrum at intermediate scales, which  is clearly visible in  Fig \ref{lab_fig1}.
Moreover, also the rapid growth in the spectrum from large towards small scales 
(scaling with a slope
 $\kappa^4$ \cite{Byrnes:2018txb}) 
is
caused by the decaying mode dynamics. See e.g. \cite{Ozsoy:2023ryl} for a review.

 With the aid of our analytical formulas, it is not difficult to analytically
 determine the position of the dip  in scenarios with large $\Pi_0$. 
 % We start from eq \eqref{resPIa}.
 We assume that the position of the dip is parametrised as $\kappa\,=\,x\,\sqrt{3/(2 \Pi_0)}$, with
 $x$ some dimensionless constant to be determined. We plug this Ansatz in  \eqref{resPIa}, and expand
 the result for large $\Pi_0$. We find the expression
 \be
 \Pi(x)\,=\,\left(1-x\right)^2+\frac{3 x^4 (6-x^2) }{10\,\Pi_0}
 +{\cal O}\left( \frac{1}{\Pi_0^2}\right)\,.
  \ee
The position of the dip in the spectrum corresponds to a choice of $x$ which makes  the first contribution in the previous
formula vanishing, leaving us with the remaining small contributions suppressed
by powers of the small quantity $(1/\Pi_0)$.  This leads to the choice  $x=1$, hence  
\be
\kappa_{\rm dip}\,=\,\sqrt{\frac{3}{2 \Pi_0}}
\hskip0.7cm;\hskip0.7cm
{\Pi}(\kappa_{\rm dip})\,=\,\frac{3}{2 \Pi_0}+{\cal O}\left( \frac{1}{\Pi_0^2} \right)\,.
\label{dippos1}
\ee
Comparing formulas \eqref{infre} and \eqref{dippos1}, we learn that the position of the dip with respect to the peak is proportional to the inverse fourth power of the total gain in magnitude of the spectrum
from large to small scales \cite{Tasinato:2020vdk}. The single parameter $\Pi_0$ hence
characterize the spectrum and all its features. 

The overall properties of the curvature spectrum we determined with the large $|\eta|$ approach
to inflation 
 are in common with several inflationary models leading to primordial black hole production, based on violations of slowroll conditions. Various other
 analytical studies based on different approaches, see e.g. \cite{Starobinsky:1992ts,Wands:1998yp,Ozsoy:2019lyy,Carrilho:2019oqg,Tasinato:2020vdk,Franciolini:2022pav,Karam:2022nym,Domenech:2023dxx}.   Our formulas have
the nice feature  of depending on a single  parameter $\Pi_0$, with a clear  physical interpretation. Moreover, the large $|\eta|$ approach is sufficiently flexible to be straightforwardly applicable to the computation of three point functions
for curvature fluctuations, leading to  novel analytical results for the bispectrum. We
explore  this topic in what comes next.

%%%%%%%%%%%%%%%%%%%%%%%%%%%%%%%%%%%%%%%%%%%%%%%%%%%%%%%%%%
\section{Large $|\eta|$ limit and non-Gaussianities}
\label{sec_bis}
%%%%%%%%%%%%%%%%%%%%%%%%%%%%%%%%%%%%%%%%%%%%%%%%%%%%%%%%%%

In the previous section, we shown how the large $|\eta|$ approach
of limit \eqref{defle} leads to a simple  analytical expression
for the power spectrum in eq \eqref{resPIa}. It
depends on a single dimensionless  parameter $\Pi_0$, and
it has   clear physical properties matching
what found in more sophisticated inflationary models of primordial
black holes. In this
section, we study the implication of eq \eqref{defle}  for the curvature
three point function and the associated bispectrum. The  topic was
briefly discussed in \cite{Tasinato:2023ukp}.  Here we  investigate it  more systematically.  
We will be able to determine fully analytical formulas for the bispectrum,  including its scale and shape dependence,  depending again on the single free
parameter $\Pi_0$ introduced in \eqref{defle}. 

It is well known that  non-Gaussianities in the statistics of fluctuations
can play an important role in the formation of primordial
black holes from inflation, for example since it affects the estimates of the threshold of PBH formation. See
e.g. \cite{Saito:2008em,Byrnes:2012yx,Young:2013oia,Bugaev:2013vba,Franciolini:2018vbk,Passaglia:2018ixg,Atal:2018neu,Biagetti:2018pjj,Atal:2019cdz,Kehagias:2019eil,DeLuca:2019qsy,Young:2019yug,Germani:2019zez,Kawasaki:2019mbl,Taoso:2021uvl,Hooshangi:2021ubn,Young:2022phe,Ferrante:2022mui,Gow:2022jfb,Tomberg:2023kli,Li:2023xtl} for works studying different aspects of this topic. In many studies, it is customary to consider 
primordial non-Gaussianity with 
a local form. In this work we analytically compute the bispectrum from
first principles using the in-in formalism, in the large $|\eta|$ limit of inflation. We show that non-Gaussian features are characterised
by a  rich shape and scale dependence.

\smallskip

Our assumptions for the behaviour of the system -- quasi de Sitter slowroll
expansion throughout the inflationary era, a part from a brief phase
of large $|\eta|$ evolution -- help us to uniquely characterise the structure
of the curvature three point function. The third order interaction Hamiltonian
for curvature perturbations in single field inflation \cite{Maldacena:2002vr} contains {\it one} term which gives
the dominant contribution to the bispectrum \cite{Kristiano:2022maq,Kristiano:2023scm}
\be
\label{expH}
{\cal H}_{\rm int}(\tau)\,=\,
-\frac12\,\int d^3 x\,a^2 (\tau)\,\epsilon(\tau)\,\eta'(\tau)\,\zeta^2(\tau,\vec x)\,\zeta'(\tau, \vec x)\,.
\ee
Since this contribution depends on the time derivative $\eta'(\tau)$, it is proportional to the large jump in the $\eta$ parameter
occurring at the beginning ($\tau=\tau_1$) and at the end ($\tau=\tau_2$)
of the non slowroll phase. Being  this epoch extremely brief in our framework, we have no option
but to consider a {\it sudden transition} among different evolution eras. We can then express the time derivative of the $\eta$
parameter, as appearing in eq \eqref{expH}, as    \cite{Kristiano:2023scm}
\bea
\label{etans1}
\eta'(\tau)  &=& \Delta \eta \left[-\delta(\tau-\tau_1)
+\delta (\tau-\tau_2) \right]\,,
\eea
the quantity $ \Delta \eta$ in eq \eqref{etans1} 
controls the jump of the $\eta$ parameter.  
 We will soon determine   its precise
value,  making use of  appropriate physical arguments.

But first,
we use the  interaction Hamiltonian \eqref{expH} to write a formal expression
for the bispectrum $B_\zeta$ at the end of inflation, $\tau=0$, defined by means of the formula:
\be
\langle \zeta_{\kappa_1} (0)\zeta_{\kappa_2} (0)\zeta_{\kappa_3}(0) \rangle
\,=\,  \delta({\vec \kappa}_1+ {\vec \kappa}_2+ {\vec \kappa}_3)\,B_\zeta(\kappa_1, \kappa_2, \kappa_3)\,.
\ee
  As usual, momentum conservation
 requires that the vectors labelling the perturbations in Fourier space
 form a closed triangle. 
  The in-in formalism prescribes to compute the three-point correlator as 
 \be
 \langle {\rm in} \Big|
{\bar T} e^{-i \int {\cal H}_{\rm int}( \tau') d  \tau'}\, \zeta_{\kappa_1}(0) \zeta_{\kappa_2} (0)\zeta_{\kappa_3}(0) \,
{ T} e^{i \int {\cal H}_{\rm int}( \tau') d  \tau'}
 \Big|
{\rm in}  \rangle\,.
\ee Using the Ansatz \eqref{etans1}, we  obtain 
the following structure 
 \cite{Kristiano:2023scm}
\bea
&&B_\zeta(\kappa_1, \kappa_2, \kappa_3)
 =
 \nonumber
 \\
 &&-2 \Delta \eta \Big(\epsilon(\tau_2) a^2(\tau_2)
 \,{\rm Im}\left[
 \left(  \zeta_{\kappa_1}(0)   \zeta^{*}_{\kappa_1} (\tau_2) \right)
  \left(  \zeta_{\kappa_2}(0)   \zeta^{*}_{\kappa_2} (\tau_2) \right)
 \left(  \zeta_{\kappa_3}(0)  \partial_{
 \tau_2} \zeta^{*}_{\kappa_3} (\tau_2) \right)
   \right]-(\tau_2\to\tau_1) \Big)
    \nonumber
 \\
 &&+{\rm perms\,.}
 \label{bisopun}
\eea
  To proceed with the computations, we now determine the value of $\Delta \eta$
%which appears as overall coefficient 
in eq \eqref{bisopun}.
\subsubsection*{The value of $\Delta \eta$, and the resulting structure of the bispectrum}
%In order to determine the overall coefficient $\Delta \eta$
%in eq \eqref{bisopun}, we proceed as follows. 
We start computing
the squeezed limit of the  expression \eqref{bisopun}. Formally it  reads
\bea
&&\lim_{q\to0} B_\zeta(\kappa, \kappa, q) \nonumber
\\
&&=\,-\,4 \Delta \eta\,\epsilon(\tau_2) a^2(\tau_2)\,|\zeta_{q}(0)|^2\,|\zeta_{\kappa}(0)|^2
\nonumber
\\
&&\times
\left\{
{\rm Im}\left[ \frac{\zeta_{\kappa}^2(0)}{|\zeta_{\kappa}(0)|^2}
\zeta_{\kappa}^*(\tau_2)(\zeta'_{\kappa}(\tau_2))^*
\right]-\frac{\epsilon(\tau_1) a^2(\tau_1)}{\epsilon(\tau_2) a^2(\tau_2)}
{\rm Im}\left[ \frac{\zeta_{\kappa}^2(0)}{|\zeta_{\kappa}(0)|^2}
\zeta_{\kappa}^*(\tau_1)(\zeta'_{\kappa}(\tau_1))^*
\right]
\right\}\,.
\nonumber
\\
\label{sqztp}
\eea
In absence of a phase of slowroll violation,  the squeezed
limit of the previous formula satisfies Maldacena consistency relation \cite{Maldacena:2002vr}: 
\bea
\lim_{q\to0 } B(\kappa, \kappa, q)%\lim_{q\to0 }\langle \zeta_{\kappa_1=\kappa}(0) %\rangle' 
&=&-\left(n_\zeta(\kappa)-1\right)\,|\zeta_{q}(0)|^2\,|\zeta_{\kappa}(0)|^2\,,
\label{conm1}
\eea
with $(n_\zeta(\kappa)-1)=d \ln \Pi(\kappa)/d \ln \kappa$. 
In our case,  we can expect the consistency relation  \eqref{conm1} to be satisfied for small
$\kappa$, in a regime $q\ll\kappa \ll 1$. In fact, modes with such small values
of momenta leave the horizon early in the inflationary phase, well before the non slowroll epoch occurs.  Correspondingly, the  decaying mode has   long time to decay: it can not
be significantly resurrected by the non slowroll phase, which occurs much later than its horizon
crossing. We can then plug the mode functions determined in Appendix \ref{appA} into eq
\eqref{sqztp}. We expand for small $\kappa$, and we impose the validity of \eqref{conm1} up to  order $\kappa^2$. This procedure determines uniquely the parameter $\Delta \eta$
in the large $|\eta|$ limit of eq \eqref{defle}. We find
\be
\label{solde}
\Delta \eta\,=\,\frac{|\eta|}{(1+\Pi_0)}+\frac{\Pi_0 (12+34 \Pi_0+25 \Pi_0^2)}{2 (1+\Pi_0)^2 (1+2 \Pi_0)}\,,
\ee
up to corrections of order $1/|\eta|$.

Armed with the precise value of $\Delta \eta$, as well as with the mode functions computed in Appendix 
\ref{appA}, we can analytically evaluate the full bispectrum using eq \eqref{bisopun}.  The overall coefficient (proportional to $\Delta \eta $) scales as $|\eta|$. But taking differences over mode functions within the parenthesis
of  \eqref{bisopun}, we find that they scale as $1/|\eta|$, 
 compensating for the overall large factor $|\eta|$.
 Taking the limit  \eqref{defle}
of large $|\eta|$ we  obtain  a finite, well defined result for the bispectrum. Its structure is
\bea
B_\zeta\left(\kappa_1, \kappa_2, \kappa_3 \right)\,=\,{\cal P}_0^2\,{\cal A}_\zeta\left(\kappa_1, \kappa_2, \kappa_3 \right)\,,
\label{bisstr1}
\eea
with ${\cal P}_0$ given in eq \eqref{defsci}, while the dimensionless function ${\cal A}$ is built
in terms of
 the combination
\be
{\cal A}_\zeta\left(\kappa_1, \kappa_2, \kappa_3 \right)\,=\,\sum_{n=1,\dots,4}\,{\cal C}_n\left(\kappa_1, \kappa_2, \kappa_3 \right)\,\Pi_0^n\,.
\label{expAA}
\ee
The analytical expression for the  functions ${\cal C}_n$ can be found  in Appendix \ref{appB}. They are  oscillatory functions of combinations of momenta. The resulting  bispectrum has a rich momentum
and shape dependence, which we explore in what follows. 
%%%%%%%
%%%%%%%
\subsubsection*{The squeezed (and not-so-squeezed) limit of the bispectrum}
%%%%%%
%%%%%%
\begin{figure}
\begin{center}
    \includegraphics[width = 0.44 \textwidth]{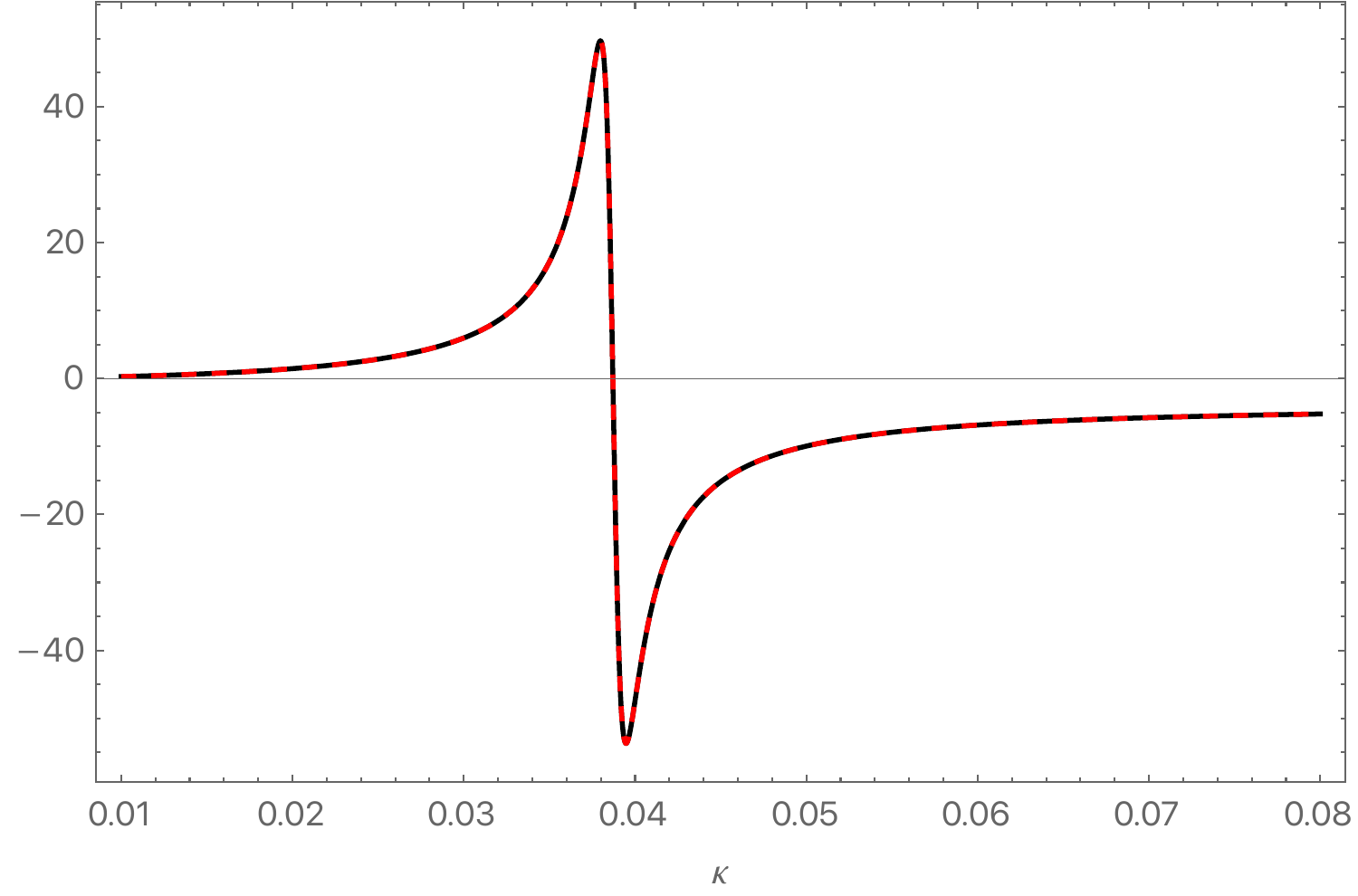}
    \includegraphics[width = 0.44 \textwidth]{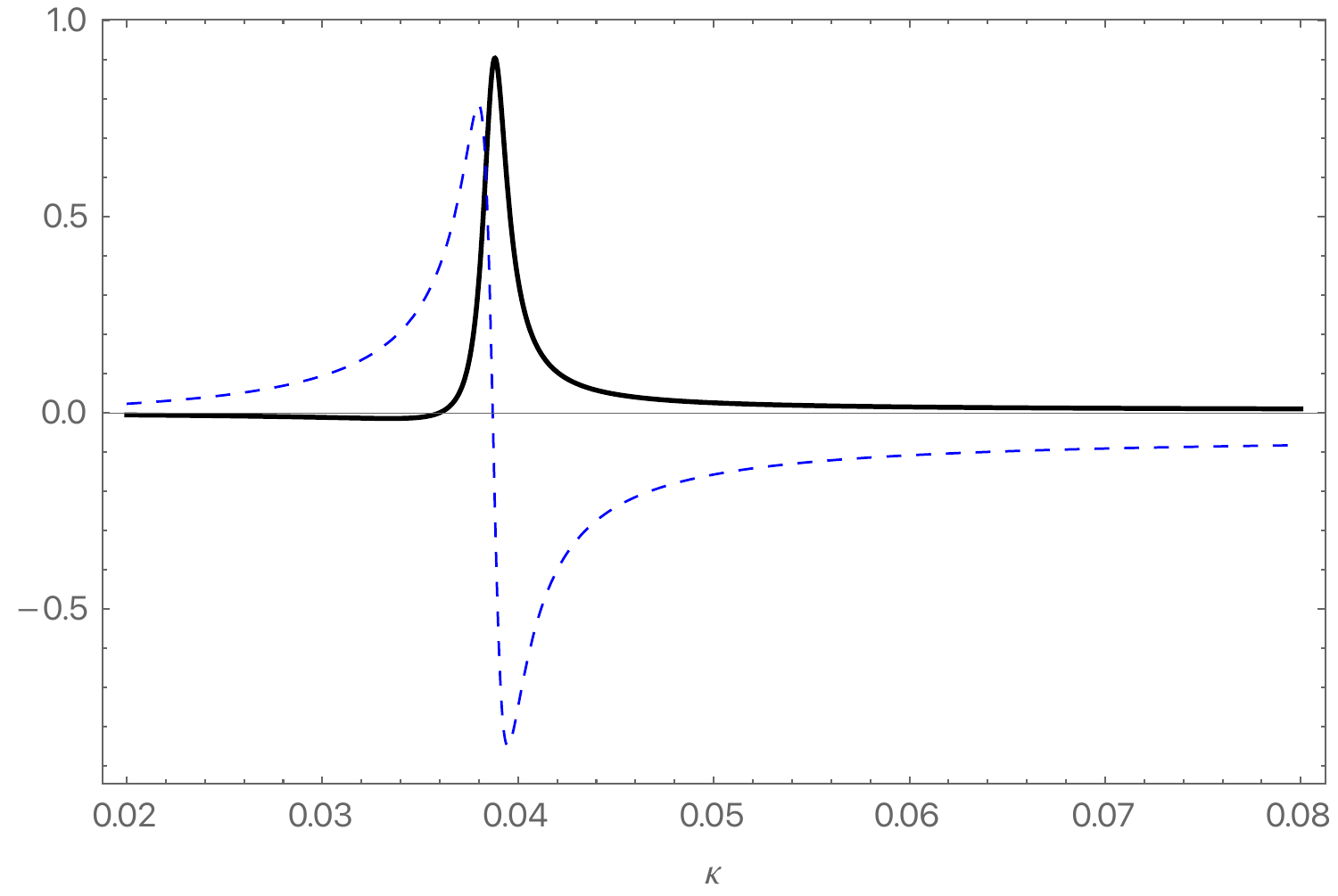}
\caption{\footnotesize{ {\bf Left}: the squeezed bispectrum satisfies Maldacena consistency relation. Red: $(1-n_\zeta)$, Dashed black the squeezed non-linear parameter $5\,f_{\rm NL}^{\rm sq}/6$
defined in eq \eqref{deffsq}.  {\bf Right}: The function $G(\kappa)$ defined in eq \eqref{NLC1},
divided by $\Pi_0^2$ (black line). 
The quantity $(ns-1)/2$ divided by $\Pi_0^{1/2}$ (dashed blue line). The normalization factors
are introduced to obtain comparable variables for the two quantities introduced. In both panels, $\Pi_0=10^3$.}}
  \label{lab_fig2}
    \end{center}
\end{figure}
The squeezed limit of the bispectrum \eqref{bisstr1} can be studied analytically.
 It satisfies Maldacena  consistency relation  
{\it for any value of $\kappa$}. Defining 
\be
\lim_{q\to0}{\cal A}\left(\kappa, \kappa, q \right)
\,=\,\frac65\, f_{\rm NL}^{\rm sq}
%-\left(n_\zeta(\kappa)-1\right)
\,\frac{\Pi(\kappa)}{\kappa^3}\frac{\Pi(q)}{q^3}\,,
\label{deffsq}
\ee
we have the relation
\be
\label{deffsq1}
f_{\rm NL}^{\rm sq}(\kappa)\,=\,\frac56\,\left(1-n_\zeta(\kappa)\right)\,.
\ee
  See Fig \ref{lab_fig2}, left panel. 
 This result might appear surprising, since it is well known
that Maldacena  consistency relation can be violated in scenarios
which include ultra slowroll phases (see e.g. \cite{Namjoo:2012aa,Martin:2012pe,Chen:2013aj}). Our result of eq \eqref{deffsq1} indicates
the 
duration \eqref{defDta} of the non slowroll era is too brief
for triggering a violation of the consistency condition -- no matter
how large $|\eta|$ is. 

As manifest in Fig \ref{lab_fig2}, left panel, 
the squeezed limit of the bispectrum is strongly scale dependent, reaching
its extrema in the region $\kappa_{\rm dip}$ around the dip in the spectrum, 
to then assume values of order one (and negative) for scale $\kappa\sim {\cal O}(1)$
towards the first peak of the spectrum. (See e.g. \cite{Chen:2005fe,Byrnes:2009pe,Byrnes:2010ft} for early works
on scale dependent non-Gaussianities.) Again, the features of $f_{\rm NL}^{\rm sq}$ are 
determined by the single parameter $\Pi_0$
controlling the spectrum.

Since we have analytical control of the bispectrum, we can also investigate 
the first order corrections ${\cal O}(q^2)$ to the squeezed bispectrum. Such contributions can be associated with the
   curvature of the background, see e.g. \cite{Creminelli:2013cga}. They 
  can manifest significant deviations from the predictions of  standard slowroll scenarios. We
 parameterise such corrections in terms of a function $G(\kappa)$:
 \be
\lim_{q\ll1}{\cal A}\left(\kappa, \sqrt{\kappa^2+q^2}, q \right)
\,=\,-\left(n_\zeta(\kappa)-1\right)\,\frac{\Pi(\kappa)}{\kappa^3}\frac{\Pi(q)}{q^3}+
q^2\,\frac{\Pi(\kappa)}{\kappa^3}\frac{\Pi(q)}{q^3}\,G(\kappa)+{\cal O}(q^3)\,.
\label{NLC1}
\ee
In computing the next-to-leading corrections \eqref{NLC1}, we focus on momenta
forming a right triangle, with one of the catheti corresponding to the small momentum $q$. 
We represent in Fig \ref{lab_fig2}, right panel,
the  function $G(\kappa)$  controlling the correction $q^2$ in eq \eqref{NLC1} (normalized by inverse powers of $\Pi_0$). As apparent from the figure, this function
has  a distinctive feature  around the  position $\kappa_{\rm dip}$ of the dip of the spectrum. This feature can
be interpreted as  due to the  decaying mode, that
  influences the  dynamics around $\kappa_{\rm dip}$. The amplitude of this
  feature is 
  parametrically larger  by a factor of order $\Pi_0^{3/2}$  with respect to the feature associated with the spectral index and $f_{\rm NL}^{\rm sq}$ (which we
  represent for comparison in Fig \ref{lab_fig2}).  This indicates 
  that the effects of the would-be decaying mode and of the non slowroll phase --  controlled by $\Pi_0$ -- become more and more accentuated
  as we leave the squeezed limit. In fact, we will now find further manifestations
  of the would-be decaying mode dynamics by studying other shapes of the bispectrum.
\begin{figure}
\begin{center}
        \includegraphics[width = 0.44 \textwidth]{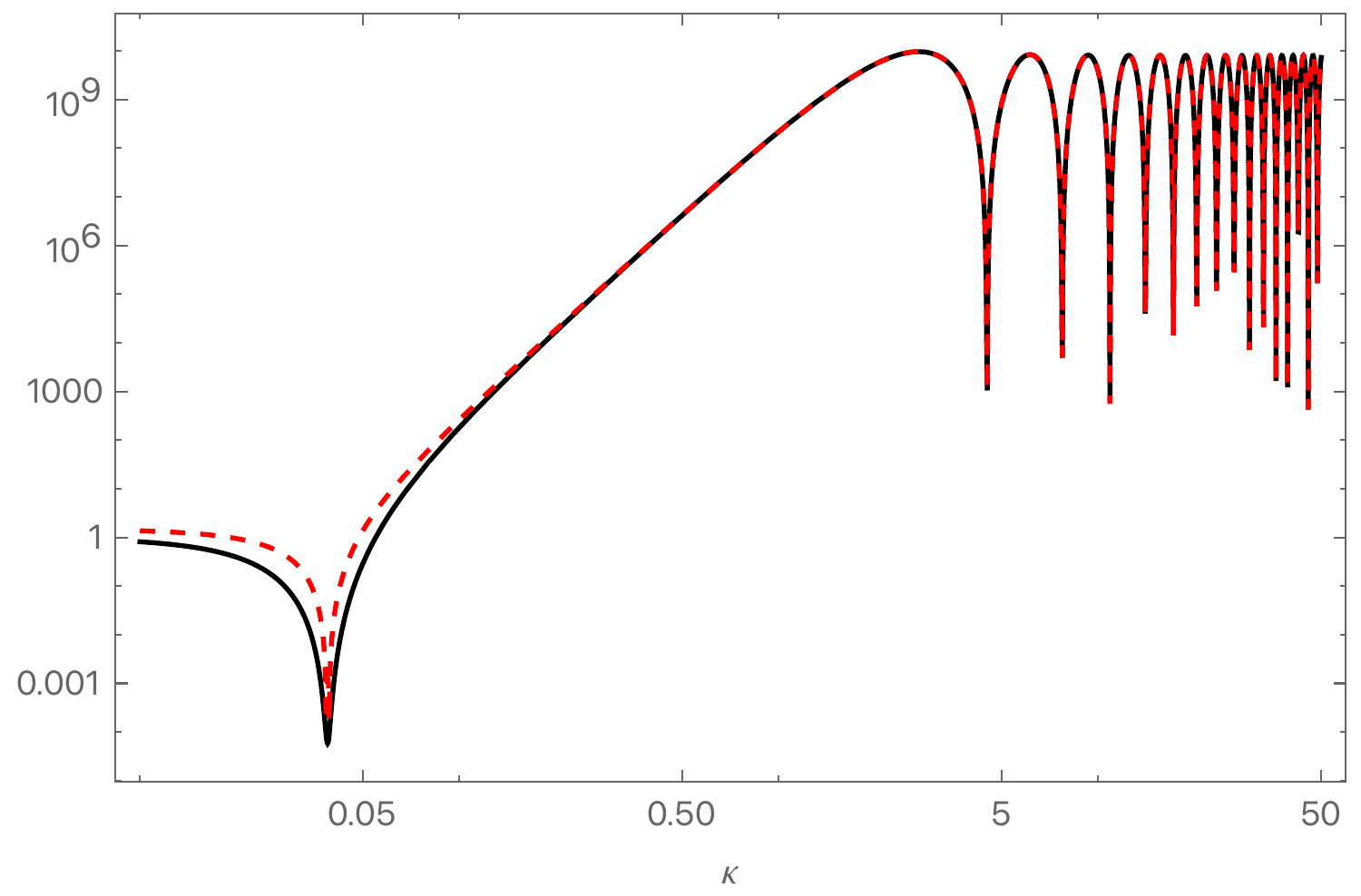}
            \includegraphics[width = 0.44 \textwidth]{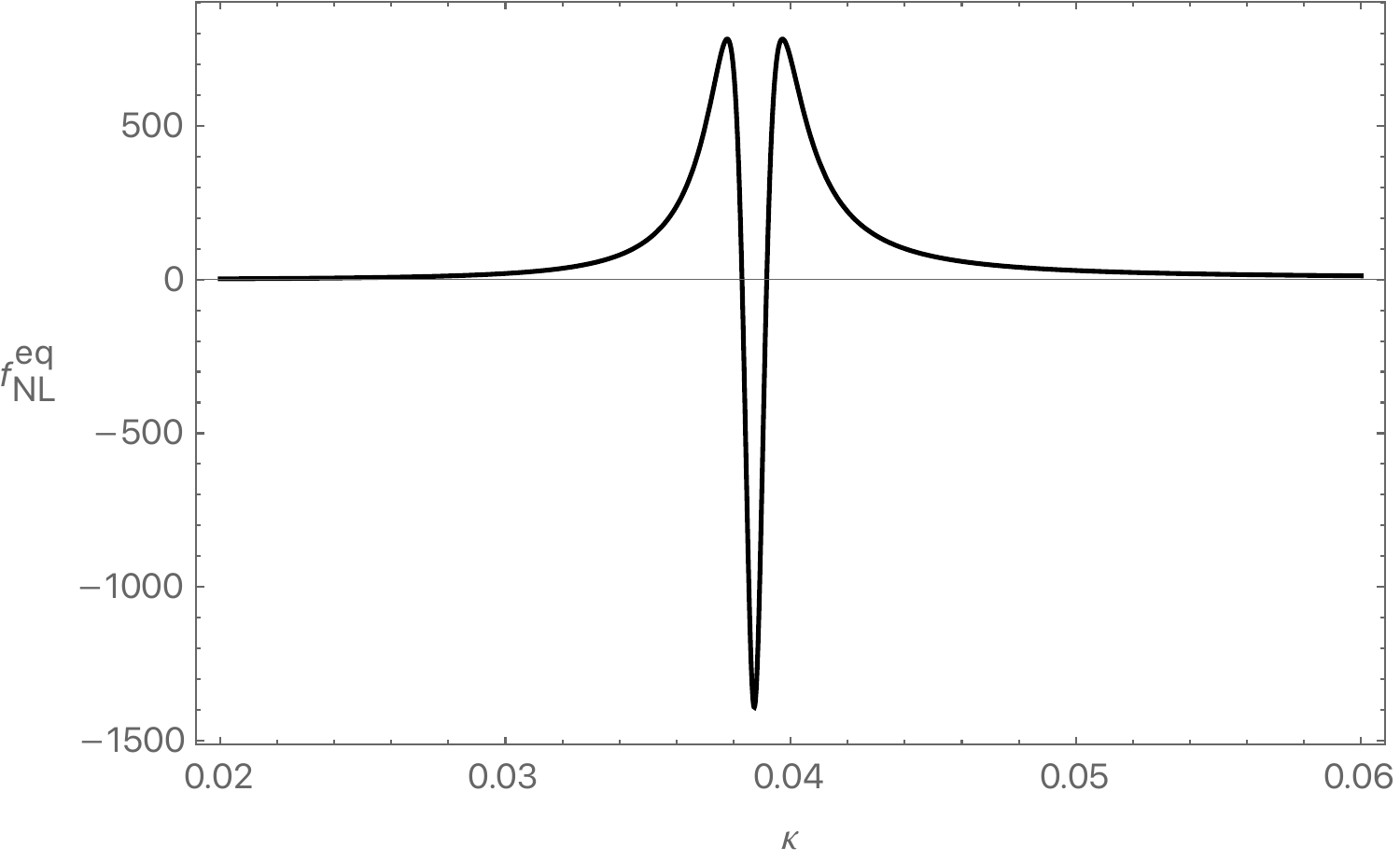}
\caption{\footnotesize{  {\bf Left:} The right hand side of eq \eqref{equaiqu}
 (black line) versus the left hand side of the same equation
(dashed red line) {\bf Right:}  The quantity  $f_{\rm NL}^{\rm eq}$ defined
in eq \eqref{deffeq}. 
In both panels, $\Pi_0=10^3$.
}}. 
  \label{lab_fig3}
    \end{center}
\end{figure}
\subsubsection*{The equilateral  shape of the bispectrum}

 For an equilateral bispectrum, where all momenta have the same size $\kappa$, we find  interesting
overlaps between the scale dependent profiles of the bispectrum and of the power spectrum. 
Recall  the expressions for the dimensionless quantities ${\cal A}$ and 
${ \Pi}$ in eqs \eqref{expAA} and \eqref{resPIa}. We  find the relation  
\be
\label{equaiqu}
\frac{\kappa^4}{6 \Pi_0}\,\big| {\cal A}_\zeta(\kappa, \kappa, \kappa)\big|\,\simeq\,\Pi^{3/2}(\kappa)\,,
\ee
is very well satisfied for the entire range of scales, see Fig \ref{lab_fig3}, left panel. 
Again,
we interpret
this behaviour as due to the dynamics of the decaying mode, that increases the magnitude
of the equilateral bispectrum in a similar  way as it does for the power spectrum. 

As customary, we define the dimensionless parameter  $f_{\rm NL}^{\rm eq}$ 
\be
\label{deffeq}
f_{\rm NL}^{\rm eq}(\kappa) \,=\,\frac{5}{18}\,\frac{B_\zeta(\kappa,\kappa,\kappa)}{P_\kappa^2}\,=\,\frac{5}{18}\,\frac{\kappa^6\,{\cal A}_\zeta(\kappa,\kappa,\kappa)}{\Pi^2(\kappa)}\,,
\ee
controlling 
the size of the equilateral bispectrum with respect to the power spectrum, and represent its scale-dependent profile in Fig \ref{lab_fig3}, right panel. The size of $f_{\rm NL}^{\rm eq}$ is relatively small
at all scales, a part from the region around $\kappa_{\rm dip}=\sqrt{3/(2 \Pi_0)}$, where
it develops pronounced features. (See also \cite{Ozsoy:2021qrg,Ozsoy:2021pws} for related computations
using a different approach based on \cite{Leach:2001zf}.)

A very similar behaviour occurs for other non-Gaussian shapes as well, and can be studied
with our analytic formula \eqref{expAA}. To conclude this section, we investigate the shape
dependence of the bispectrum using a convenient graphical representation
in terms of triangular plots. 

\subsubsection*{Graphical representation of more general shapes of the bispectrum}
\begin{figure}
\begin{center}
    \includegraphics[width = 0.4 \textwidth]{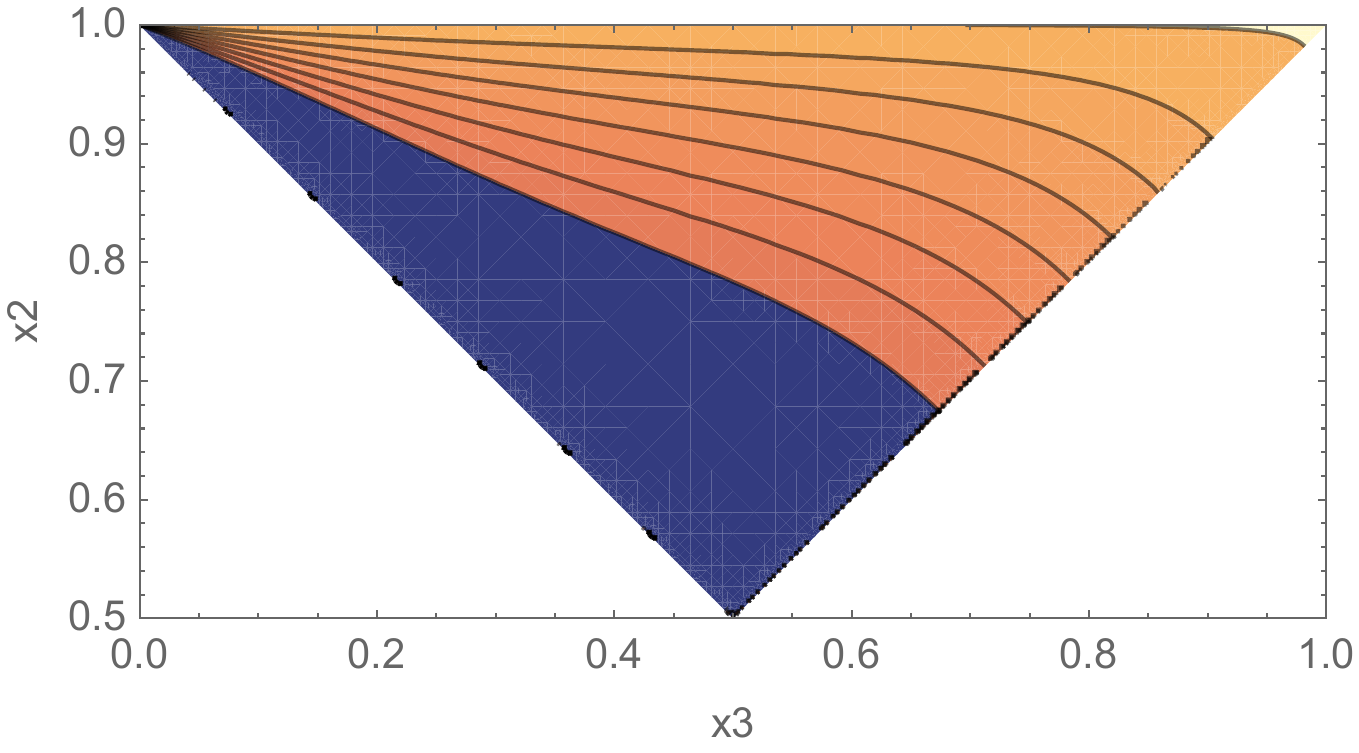}
        \includegraphics[width = 0.07 \textwidth]{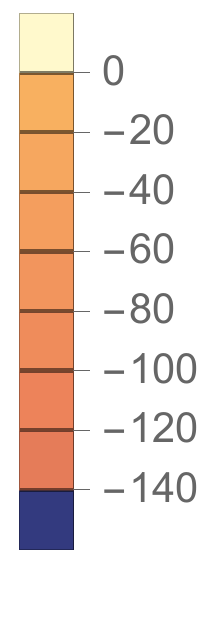}
          \includegraphics[width = 0.4 \textwidth]{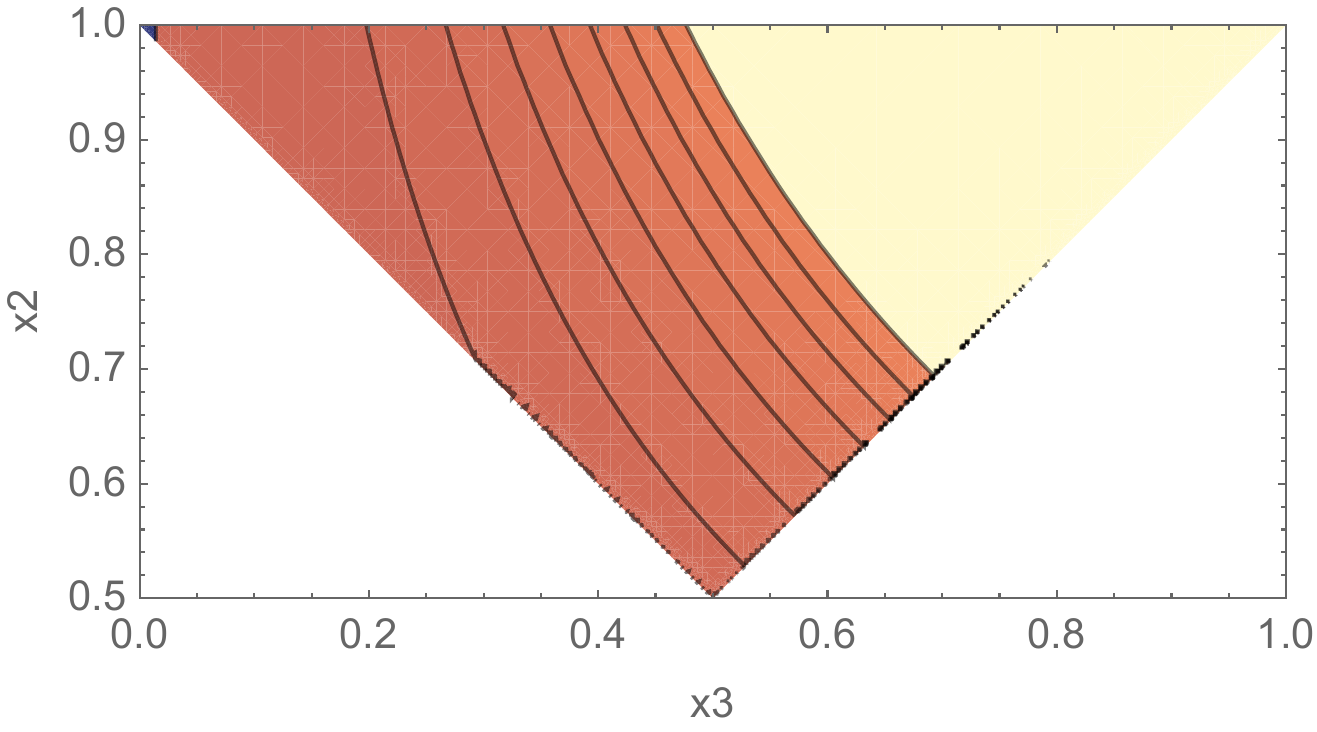}
        \includegraphics[width = 0.07 \textwidth]{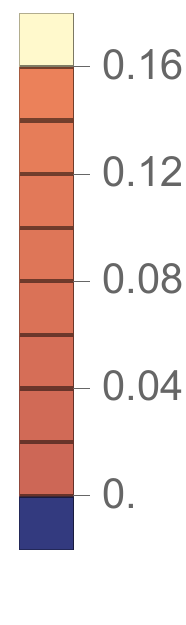}
\caption{\footnotesize 
Triangular plots of the function $S(\kappa, x_2, x_3)$ of eq \eqref{deffS}, represented
with the criteria discussed in the text. 
{\bf Left} $S(\kappa_{\rm dip}, x_2, x_3)$.  {\bf Right} $S(1, x_2, x_3)$.}
  \label{lab_fig4}
    \end{center}
\end{figure}

For representing the shape dependence of the bispectrum as a function
of the scale,  we use the graphical device introduced in \cite{Babich:2004gb} (see
also \cite{Baumann:2009ds} for a pedagogical review). We 
define
the function
\be
\label{deffS}
S(\kappa_1, \kappa_2, \kappa_3)\,=\,N \,\kappa_1^2\,\kappa_2^2\,\kappa_3^2\,
B_\zeta(\kappa_1, \kappa_2, \kappa_3)\,,
\ee
where the normalization constant $N$ is selected such that $S(1, 1, 1)\,=\,1$. 
We then introduce the coordinates $x_2\,=\,\kappa_2/\kappa_1$, $x_3\,=\,\kappa_3/\kappa_1$, and represent the magnitude of  $S(\kappa, x_2, x_3)$ in triangular plots as
in Fig \ref{lab_fig4}. Ordering momenta such that $x_3\le x_2\le 1$, the triangular
inequality requires $x_2+x_3>1$. 
To avoid
representing configurations which are equivalent, we only focus on the region 
$1-x_2\le x_3\le x_2$.
Given the fact that our  bispectrum is strongly dependent
on the scale, the result depends on the choice of the
first argument $\kappa$ in the quantity $S(\kappa, x_2, x_3)$.  We make
the two choices $\kappa=\kappa_{\rm dip}$ and $\kappa=1$. While for 
 $\kappa=1$ the bispectrum is mostly enhanced for the equilateral shape, the behaviour 
 at  $\kappa_{\rm dip}$ is more rich, and the spectrum is enhanced for elongated triangles, where the size of two of the momenta is well larger than the third one. Hence, depending on the scales one
 considered, different shapes can become dominant, with distinct consequences
 for observables controlling the physics of PBH.

In fact, is there any specific physical implication for the rich scale dependence of the bispectrum
in our system, as
we explored so far? In the next section,  we will study some consequences
of our findings for the statistics of scalar-induced gravitational waves.

%%%%%%%%%%%%%%%%
\section{Gravitational waves and non-Gaussianity}
\label{sec_gw}
%%%%%%%%%%%%%%%

In the investigations we carried on so far, we demonstrated that the analytical
bispectrum of eq \eqref{expAA} has a rich shape and scale dependence,    more complex 
than a pure local-type  non-Gaussian statistics. The information we obtained
can be important when studying observables that involve convolution integrals
over all scales. Such quantities can in fact be sensitive to the large scale
features of the bispectrum we investigated in section \ref{sec_bis}. 

Explicit examples are observables built in terms of tensor modes sourced {\it at second
order} by  scalar fluctuations which are enhanced at small scales, for example
by  non slowroll epochs as  considered in sections \ref{sec_ps} and \ref{sec_bis}.
 See e.g. \cite{Matarrese:1992rp,Matarrese:1993zf,Ananda:2006af,Baumann:2007zm,Saito:2008jc,Saito:2009jt,Espinosa:2018eve,Domenech:2019quo} for some of the original papers on the subject,  \cite{DeLuca:2019ufz,Inomata:2019yww} for discussions on
 gauge issues on this framework,
   and \cite{Domenech:2021ztg} (and references therein) for a comprehensive review.  Integrations
involving convolutions depend on the power spectrum and bispectrum at all scales.
 They are  sensitive to the features of the statistics of fluctuations
that we studied in the previous section. In fact, observable relative to the induced tensor two and three point functions constitute an important, although indirect,  experimental window on small-scale scalar fluctuations. 
We proceed to study  this phenomenon explicitly.

Schematically,
scalar perturbations with an enhanced curvature spectrum $P_{\zeta}(k)$  (recall its definition in eq \eqref{gedeP}) can source a tensor spectrum  $P_{h}$
at second order in fluctuations
\be\label{schem1}
P_h (k)\,=\,\int d k'\,f(k, k')\,P_\zeta(k')\,P_\zeta(k-k')\,,
\ee
where $f(k, k')$ is a kernel
function depending on cosmology. For convenience, in this section we 
make use of dimensionful momenta $k$: their dimensionless version $\kappa$
can be easily obtained in terms of the definition \eqref{defkap}.
 Expressions  as \eqref{schem1}
are derived under the hypothesis of Gaussian primordial curvature fluctuations. Local type  non-Gaussianity can modulate the scale dependence of the induced
tensor spectrum, with interesting phenomenological implications \cite{Cai:2018dig,Unal:2018yaa,Yuan:2020iwf,Atal:2021jyo,Adshead:2021hnm,Chang:2022nzu,Garcia-Saenz:2022tzu,Li:2023qua}. Here
we explore the implications of an enhanced curvature bispectrum for another interesting observable: the tensor-tensor-scalar bispectrum $B_{TTS}$ evaluated 
during radiation domination. The quantity  $B_{TTS}$ involves
two (scalar induced) tensor modes $h$ and one Bardeen scalar mode $\Phi$.  Schematically, in the squeezed limit we expect a relation like the following one (soon we will be more precise)
\be
\label{schem2}
B_{TTS}(k_1, k_2, k_3)\,=\,\int d  q\,\tilde{f}(k_1, k_2,q )\,B_{\zeta}(k_1, k_2, q)\,P_\zeta(k_1-q)\,,
%\,P_\zeta(\kappa')
\ee
for $\tilde f$ some  kernel
function.  A squeezed version of the bispectrum  of eq \eqref{schem2}
can  correlate
(scalar induced) stochastic backgrounds   probed
by gravitational wave experiments at small scales, with a large scale scalar modes  probed by cosmic microwave background observations. It is the
 observable investigated in \cite{Adshead:2020bji} (see also \cite{Malhotra:2022ply,Dimastrogiovanni:2022eir,LISACosmologyWorkingGroup:2022jok,Dimastrogiovanni:2022afr,Dimastrogiovanni:2021mfs,Braglia:2021fxn,Ricciardone:2021kel,Malhotra:2020ket,Orlando:2022rih}) for carrying on a `multimessenger cosmology' program. It suggests new experimental avenues for probing
 the physics of early universe, and its detection and characterization would provide
a smoking gun for the  
  primordial origin of the gravitational stochastic background. Besides
  the squeezed limit, the bispectrum  $B_{TTS}$ might be
  phenomenologically relevant for other elongated shapes also, above all
  if its size is large
  % therefore it is worth investigating its properties.
 
 \smallskip
 
 We leave these interesting phenomenological applications aside, and we 
 proceed with computing more precisely the observable $B_{TTS}$ described
 above eq \eqref{schem2}.
The actual computations are rather technical, and we relegate them to 
Appendix \ref{appC}. Let us start discussing the results 
  for the
squeezed limit, where expressions simplify. We find  ($k_1=k_2=k$)
\bea
\label{gwsq2}
\lim_{k_3\to0} B_{TTS}(k,k,k_3)
\,=\,\frac{12}{5}\,
f_{\rm NL}^{TTS}(k)\,P_{h}(k)
P_{\zeta}(k_3)\,,
 \eea
where
\bea
\label{defttsf}
f_{\rm NL}^{TTS}(\tau, k)\equiv \frac{5}{12}\,\frac{\int d^3 q \,{{ P}_\zeta( |\vec k-\vec q|)}%{|\vec k_1-\vec q_1|^3}
{{ P}_\zeta( q)}\,\,\left[1-n_\zeta(q)\right]\,{\cal F}_0(\tau,\vec k,\vec q)}{\int d^3 q \,{{ P}_\zeta( |\vec k-\vec q|)}%{|\vec k_1-\vec q_1|^3}
{{ P}_\zeta( q)}\,{\cal F}_0(\tau,\vec k,\vec q)}\,,
\eea
while the tensor spectrum is
\be
P_h(\tau, k)\,=\,\int d^3 q \,{{ P}_\zeta( |\vec k-\vec q|)}%{|\vec k_1-\vec q_1|^3}
{{ P}_\zeta( q)}\,{\cal F}_0(\tau,\vec k,\vec q)\,.
\label{defph}
\ee
The explicit
expression for the kernel function ${\cal F}_0$ can be found in eq \eqref{defFz}.
In writing eq \eqref{defttsf} 
 we use the fact that the squeezed
limit of the scalar bispectrum satisfies Maldacena consistency relation, see section \ref{sec_bis}. 

 \smallskip
\begin{figure}
    \includegraphics[width = 0.45 \textwidth]{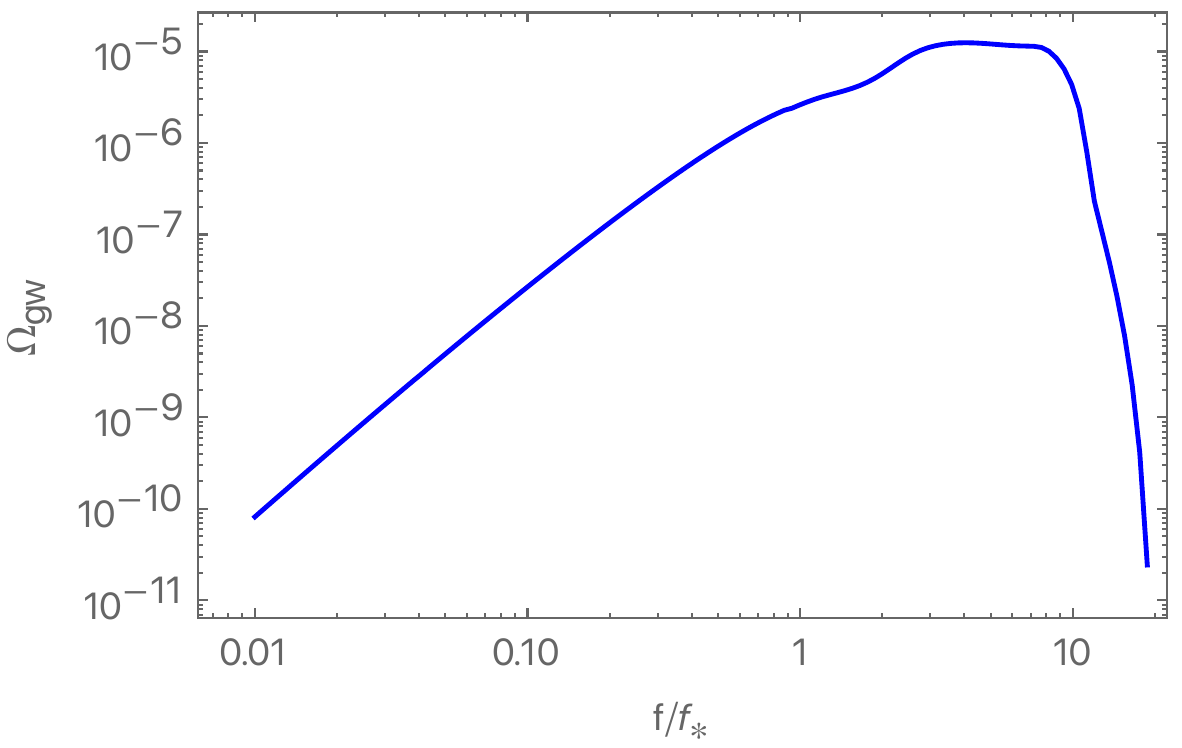}
          \includegraphics[width = 0.45 \textwidth]{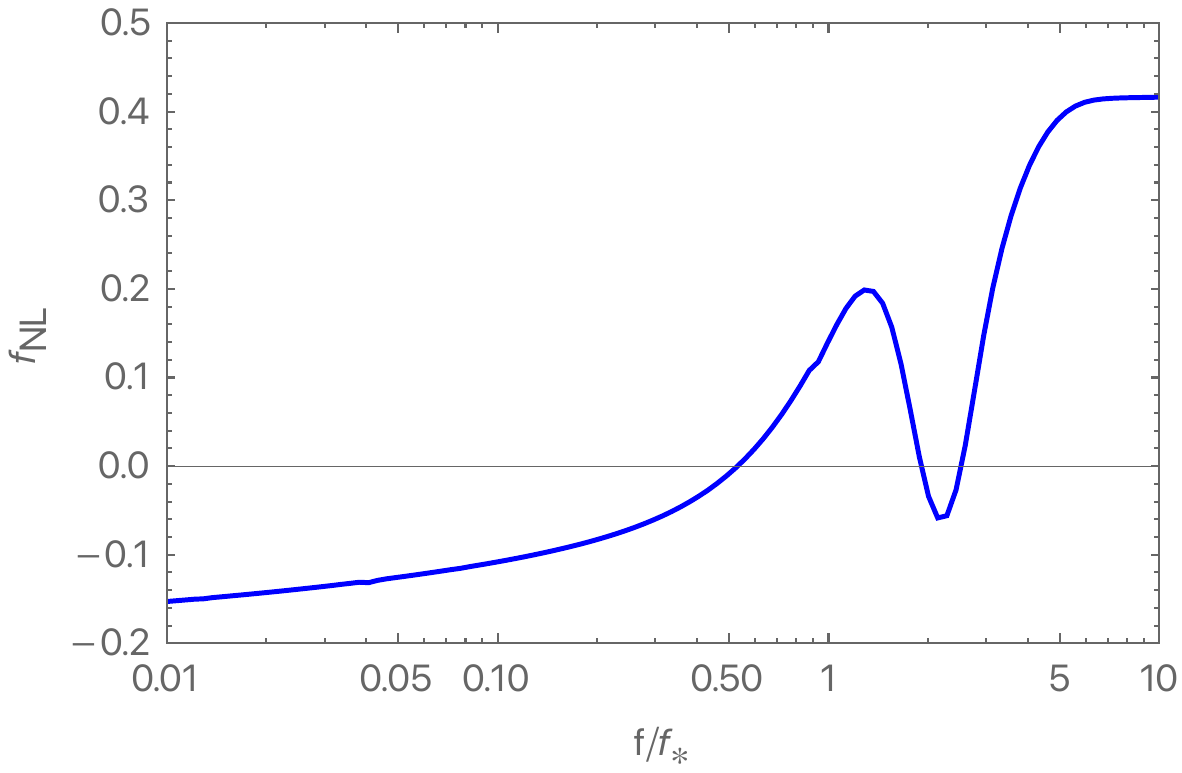}
\caption{\footnotesize {\bf Left:} plot of the gravitational wave
energy density induced by a  scalar spectrum derived from eq \eqref{defph}, with $\Pi_0=10^3$, under
the conditions discussed in the main text. {\bf Right:} plot of the non-linear parameter 
$f_{\rm NL}^{TTS}$ in eq \eqref{defttsf} computed with the same curvature  spectrum of the left figure.}
  \label{fig2bbC}
\end{figure}
Starting from eq \eqref{defph} it is straightforward  to derive the expression for the gravitational
wave energy density $\Omega_{\rm GW}$, and perform the integrals numerically by implementing the convenient formulas
of \cite{Kohri:2018awv}. We represent the result in Fig \ref{fig2bbC}, left panel, where we compute the $\Omega_{\rm GW}$ induced by the analytical scalar spectrum of  eq \eqref{bisstr1}. 
 To avoid numerical
issues, we smoothed out the oscillatory features occurring at small scales $\kappa>1$, and
we truncated the power at $\kappa=10$  (i.e. we set $\Pi(\kappa)=0$ for $\kappa>10$). We converted to frequency $f=2 \pi k$.
% and we have chosen a reference frequency $$
The result is a broad gravitational spectrum, with an ample plateau for 
$1\le f/f_* \le 10$,
with $f_*$ a reference frequency.  See  Fig \ref{fig2bbC}, left panel. In Fig \ref{fig2bbC}, right panel, we represent
the non linear parameter $f_{\rm NL}^{TTS}$ introduced in eq \eqref{defttsf}, computed
starting from the curvature power spectrum as described above. $f_{\rm NL}^{TTS}$
 has a strong dependence on the scale, and its typical magnitude is not large, being    of the order  of $10^{-1}$. It has an oscillatory behaviour of $f_{\rm NL}^{TTS}$
at frequencies  around the peak of the tensor spectrum. We interpret it as a consequence  of the  sign
change and amplified features of the scalar spectral index $n_\zeta(\kappa)-1$ appearing
in the integrand of eq \eqref{defttsf} (see Fig \ref{lab_fig2}), which are amplified where the tensor spectrum is most
enhanced. 
It would be interesting to explore  whether other  shapes 
%computable
%with the formulas given in Appendix \ref{appC}, 
 can reach larger
magnitudes in this set-up. 

In fact, the general structure of the $B_{TTS}$ bispectrum is given in eq \eqref{totbisgw}, which we report here:
\bea
B_{TTS}(\vec k_1,\vec k_2,\vec k_3)&=&
\int d^3 q \,{{ P}_\zeta( |\vec k_1-\vec q|)}%{|\vec k_1-\vec q_1|^3}
\, B_\zeta\left( q,|\vec k_3+\vec q| ,  k_3
\right)\,{\cal F}_1(\tau,\vec k_1, \vec k_3, \vec q)
\nonumber
\\
&+&{{ P}_\zeta(  k_3)} \int d^3 q \,%{|\vec k_1-\vec q_1|^3}
 \,B_\zeta\left( |\vec k_1+\vec k_3|,|\vec q+ \vec k_1+\vec k_3|, q
\right)\,{\cal F}_2(\tau,\vec k_1, \vec k_3, \vec q)\,,
\label{totbisgw2}
\eea
where the quantities ${\cal F}_{1,2}$ can be found in eqs 
\eqref{defbf1}
and 
\eqref{defbf2}. Such  expression shows that the scale and shape dependence
of the bispectrum can play an important role. For example,  going beyond the squeezed limit (considering the leading $k_3^2$ corrections to eq \eqref{gwsq2},
which might still be relevant for coupling large and small
scale modes) the convolution integral
in the second line in eq \eqref{totbisgw2} starts to contribute to the result, in a way that depends
on shapes of the bispectrum  different from the local one. We learned in 
section \ref{sec_bis} how the effects of the non slowroll phase, and of a large
parameter $\Pi_0$,  can considerably influence the amplitude of the bispectrum
for shapes different from local. It will be important to further study the behaviour of
$B_{TTS}$ for general shapes and its phenomenological consequences.

\smallskip
While in this work we focussed  on the bispectrum, the large $|\eta|$ approach can also be applied to compute  the trispectrum at leading order in $1/|\eta|$. We are hopeful that, starting
from the fourth order interaction
Hamiltonian, and proceeding as in section \ref{sec_bis}, a full analytic expression for the scalar trispectrum
can be determined. We expect such expression to exhibit a rich shape and scale dependence, as we found for the bispectrum. 
Such result would be  relevant for the computation of the induced
spectrum of gravitational waves, that through convolution integrals is modulated by the trispectrum \cite{Adshead:2021hnm,Garcia-Saenz:2022tzu,Maity:2023qzw}. While most of works so far focussed on local type non-Gaussian  correlators, it would be interesting to extend computations to the case of a more complex trispectrum derived from first principle calculations in the $1/|\eta|$ limit.

\subsection*{Acknowledgments}
It is a pleasure to thank Jacopo Fumagalli for discussions and Ameek Malhotra for feedback. 
GT is 
 partially funded by the STFC grants ST/T000813/1
and ST/X000648/1. For the purpose of open access, the author has applied a Creative Commons Attribution licence to any Author Accepted Manuscript version arising.

%%%%%%%%%%%%%%%%%%%%%%%%%
\begin{appendix}
%%%%%%%%%%%%%%%%%%%%%%%%

%%%%%%%%%%%%%%%%%%%%
\section{The mode functions during the brief non slowroll phase}
\label{appA}
%%%%%%%%%%%%%%%%%%%%%
In this Appendix we briefly review the results of \cite{Tasinato:2020vdk} which analytically compute
the solution for mode functions in a scenario of inflation undergoing a brief, but
drastic phase of non slowroll evolution. We parameterize the background
 evolution as constituted by three phases:
 \begin{enumerate}
 \item 
  An initial slowroll phase, where
 the slowroll parameters $\epsilon$ and $\eta$ are very small and
 the evolution can be approximated as de Sitter phase. The constant value of the
 parameter $\epsilon$ is denoted with $\epsilon_1$.
 \item
  A second, brief
 non slowroll epoch lasting
 between two conformal times $\tau_1$ and $\tau_2$. The parameter $\eta$ is negative and large in absolute magnitude,  while the small, time dependent parameter $\epsilon(\tau)$
 rapidly decreases  (and the Hubble parameter remains  almost constant). 
 \item
   Finally a third
 era of slowroll evolution for $\tau_2\le\tau\le0$,  where  $\epsilon$ and $\eta$
 are both small and constant.   The  value of the
 parameter $\epsilon$ in this phase is denoted with $\epsilon_2$.
 \end{enumerate}
 We rescale the curvature perturbation $\zeta_k(\tau)$ as
 \be
 \varphi_k(\tau)\,=\,z(\tau)\,{\zeta_k (\tau)}\,.
 \ee
 The Mukhanov-Sasaki equation  for the variable $\varphi_k(\tau)$ reads
% reads %(we use in this appendix the physical momentum $k$, differently from )
 \be
 \label{MS1}
  \varphi_k''(\tau)+\left(k^2-\frac{z''(\tau)}{z(\tau)} \right)\,\varphi_k(\tau)=\,0\,.
 \ee
 The general expression of the pump field $z(\tau)$ is
 \be
 z(\tau)\,=\,a(\tau) \sqrt{2 \epsilon(\tau)}\,.
 \ee
 Given the conditions described above, in a quasi de Sitter limit
 the pump field
 can be paramaterised
as follows
\be
z(\tau)\,=\,
\begin{cases}-\frac{\sqrt{2 \,\epsilon_1}}{H_0\,\tau} & {\text{for $\tau\le \tau_1$}}
\\ \\-\frac{\sqrt{2 \,\epsilon(\tau)}}{H_0\,\tau} & {\text{for $\tau_1\le \tau\le \tau_2$}}
\\ \\ -\frac{\sqrt{2 \,\epsilon_2}}{H_0\,\tau} & {\text{for $\tau_2\le \tau \le 0$}}
\end{cases}
\ee  
with $\epsilon(\tau)$ a continuous function, and  $H_0$
the (nearly constant) Hubble parameter during inflation. We define the  parameter $\eta$
as
\be
\eta\,=\,-\lim_{\tau\to\tau_1^+}\,\frac{d \ln \epsilon(\tau)}{d \ln \tau}\,,
\ee
and we consider it constant, negative, and large in absolute value. 

We parameterize the solution of Mukhanov-Sasaki equation \eqref{MS1} in terms of the following formal
series
\be
\label{ans1}
\varphi_k(\tau)\,=\,-i\frac{H_0\,e^{-i k \tau}}{2\sqrt{\epsilon_1\,k^3}}\,z(\tau)\,
\left[1 +i k \tau+ (i k \tau_1)^2 A_{(2)}(\tau)+ (i k \tau_1)^3 A_{(3)}(\tau)+\dots \right]\,,
\ee
where the functions $A_{(n)}(\tau)$, $n\ge2$, have to be determined. Plugging
the Ansatz  \eqref{ans1} into the Mukhanov Sasaki eq \eqref{MS1}, we find the following
coupled system of equations for the quantities $A_n(\tau)$:
\bea
\label{formeq1}
\left[\frac{\epsilon(\tau)}{\tau^2}\,\tau_1^2\,A_{(2)}'(\tau) \right]'
&=&\frac{\epsilon'(\tau)}{\tau}\,,
\\
\label{formeq2}
\left[\frac{\epsilon(\tau)}{\tau^2}\,
\left( 
\tau_1 \,A_{(n)}'(\tau) -A_{(n-1)}(\tau)\right) \right]'
&=&\frac{\epsilon(\tau)\,A_{(n-1)}'(\tau) }{\tau^2} \hskip0.7cm, {\text{for $n>2$}}\,.
\eea
If $\epsilon(\tau)$ is constant, then a consistent solution is $A_{(n)}=0$ for $n\ge2$, and eq \eqref{ans1}
reduces to the usual solution for mode functions in the de Sitter limit. But we are interested on evaluating the effects a time dependent $\epsilon(\tau)$ (and a large $\eta$ parameter) during the non slowroll phase of evolution. A formal solution for eq \eqref{formeq1} is
\be
\label{formsol1}
\tau_1^2\,A_{(2)}(\tau)\,=\,\int_{-\infty}^\tau\,d \tau_a\,\frac{\tau_a^2}{\epsilon(\tau_a)}
\,\left(
\int_{-\infty}^{\tau_a}\,d \tau_b\,\frac{\epsilon'(\tau_b)}{\tau_b}
\right)\,.
\ee
For $\tau<\tau_1$, the $\epsilon$ parameter is constant and $A_{(2)}(\tau)$ vanishes as desired: it acquires a non-zero value starting from $\tau=\tau_1$ onwards. 

Considering values of $A_{(n)}(\tau)$ for $n>2$, we find the following formal solution
for eq \eqref{formeq2}:
\be
\label{formsol2}
\tau_1\,A_{(n)}(\tau)\,=\,\int_{-\infty}^{\tau} d \tau_a\,A_{(n-1)}(\tau_a)
+\int_{-\infty}^{\tau}\,d \tau_a\,\frac{\tau_a^2}{\epsilon(\tau_a)}
\left( 
\int_{-\infty}^{\tau_a}\,d \tau_b\,\frac{\epsilon(\tau_b)\,A_{(n-1)}'(\tau_b)}{\tau_b^2}
\right)\,.
\ee
Interestingly, we can now exploit   the fact that the duration of the non slowroll phase
is small, $-(\tau_2-\tau_1)/\tau_1\ll1$. We expand in Taylor series
 the  previous formal solutions, focussing on the leading term (more details
 in \cite{Tasinato:2020vdk}). 
For $A_{(2)}$ we find, when evaluated at $\tau_1 \le \tau \le \tau_2$:
\bea
A_{(2)}(\tau)&=&\left(\frac{d \ln \epsilon(\tau)}{d \ln \tau}
\right)_{\tau=\tau_1}\,\frac{\left(\tau -\tau_1 \right)^2}{\tau_1^2}+{\cal O}\left[(\tau-\tau_1)^3 \right]\,,
\nonumber
\\
&=&-\eta\,\frac12 \,\frac{\left(\tau -\tau_1 \right)^2}{\tau_1^2}+{\cal O}\left[(\tau-\tau_1)^3 \right]\,.
\eea
Proceeding analogously and Taylor expanding for higher $n$, we find the following leading contributions
\bea
\label{tay1}
A_{(n)}(\tau)&=&-\eta\,\frac{2^{n-2}}{n!}\,\frac{\left(\tau -\tau_1 \right)^n}{\tau_1^n}\,, \hskip0.7cm {\text{for $n\ge2$}}\,.
\eea
Plugging the solutions of eq \eqref{tay1} into eq \eqref{ans1}, one finds an exponential series, which
can be  resummed giving 
\be
\varphi_k(\tau)\,=\,-i\frac{H_0\,e^{-i k \tau}}{2\sqrt{\epsilon_1\,k^3}}\,z(\tau)\,
\left[1 +i k \tau+
\frac{\eta}{4} \left( 1+2 i k \left( \tau-\tau_1\right)-e^{2 i k (\tau-\tau_1)}
\right)\right]\,
\hskip0.5cm {\text{for $\tau_1\le\tau\le\tau_2$}}\,.
\ee
This expression, valid for $\tau\le \tau_2$, can then be matched to a solution of Mukhanov-Sasaki
equation in the interval $\tau_2\le \tau\le0$. Recall that in this epoch we return to a slowroll regime 
where $\epsilon$ and $\eta$ are negligibly small, hence the most general solution for the mode function is
\be
\label{finm2}
\varphi_k(\tau)\,=\,-i\,{\cal K}_1\frac{H_0\,e^{-i k \tau}}{2\sqrt{\epsilon_1\,k^3}}\,z(\tau)\,
\left(1 +i k \tau\right)-i\,{\cal K}_2\frac{H_0\,e^{i k \tau}}{2\sqrt{\epsilon_1\,k^3}}\,z(\tau)\,
\left(1 -i k \tau\right)\,.
\ee
The scale dependent quantities ${\cal K}_{1,2}$ are determined by Israel matching 
the expression \eqref{finm2} with 
the solution \eqref{ans1} at time $\tau_2$.  Defining $\Delta \tau=-(\tau_2-\tau_1)/\tau_1$,
the result is 
\bea
\label{expb1}
{\cal K}_1&=&1-\frac{\eta}{8(1-\Delta \tau)^2\,k^2\,\tau_1^2}
\left[ 1-e^{-2 i k \tau_1 \,\Delta \tau }
+ k \tau_1 \,\Delta \tau
\left( i + 2 k \tau_1(1-\Delta \tau)\right)
  \right]\,,
\\
{\cal K}_2&=&
\frac{\eta e^{-2 i k \tau_1 \,(1-\Delta \tau )}}{8(1-\Delta \tau)^2\,k^2\,\tau_1^2}
\left[ 1
+2 i k \tau_1
-e^{-2 i k \tau_1 \,\Delta \tau }
%+ k \tau_1 \,\Delta \tau
\left( i + 2 k \tau_1(1-\Delta \tau)\right)
  \right]\,.
\label{expb2}
\eea
It is straightforward to compute the two point function for curvature
perturbations at the end of inflation $\tau=0$ in terms of $|\varphi_k(\tau)|^2$ using
eqs \eqref{finm2} and \eqref{expb1}, \eqref{expb2}. The result can be used to prove eq \eqref{infre} in
the main text (see also \cite{Tasinato:2020vdk}). Taking the large $|\eta|$ limit
following eq \eqref{defle}, one recovers
the
expression \eqref{solzm} in the main text (expressed
in terms of the dimensionless momenta $\kappa$ defined in eq \eqref{defkap}).

%%%%%%%%%%%%%%%%%%%%%%
\section{Explicit expressions for the  coefficients ${\cal C}_n$}
\label{appB}
In this Appendix we collect the explicit  functions ${\cal C}_n$, $n=1,\dots4$, which
form the  analytic scalar bispectrum of eq \eqref{bisstr1}. 
After defining
 \bea
\Delta_{pq}&=&1+\delta_{pq}
\hskip0.6cm,\hskip0.6cm
\Delta_{pqr}\,=\,\left(1+\delta_{pq}\right)\left(1+\delta_{qr}+\delta_{pr}\right)
\eea
we  make use of the convenient
combinations
%%%%%%%%%%%%%%%%%%%%%%
% combinations (Ferguson Shellard 0812.3413):
\bea
K_p&=&\sum_i \,k_i^p\,,
\\
K_{pq}&=&\frac{1}{\Delta_{pq}}\sum_{i\neq j} \,k_i^p k_j^q\,,
\\
K_{pqr}&=&\frac{1}{\Delta_{pqr}}\sum_{i\neq j\neq \ell} \,k_i^p k_j^q k_\ell^r\,,
\eea 
introduced in \cite{Fergusson:2008ra}. The previous combinations
allow us to express the functions ${\cal C}_n$ as
%and
\bea
{\cal C}_1&=&\frac{\left(K_{114}+3 K_{123}+6 K_{222}-K_4-2 K_{13}-2 K_{22}-2 K_{112}\right)}{K_{333}}\,j_1(\kappa_1+\kappa_2+\kappa_3)\nonumber
\\
&&+\frac{\left(K_{14}+3 K_{23}+2 K_{113}+2 K_{112}\right)}{K_{333}}\,j_0(\kappa_1+\kappa_2+\kappa_3)
\\
{\cal C}_2&=&\frac{1}{K_{444}}\Big[10 K_{134}+14 K_{233}+9 K_{224}+2K_{125}   
-
K_{112}-K_{13}
\nonumber\\
&&
- K_{24}-3 K_{222}-K_{15}-2 K_{123}-3 K_{114}\Big]\,
 j_0(\kappa_1+\kappa_2+\kappa_3)
\nonumber
\\
&&
+\frac{1}{K_{444}}\Big[K_{225}+7 K_{234}+18 K_{333}
+3 K_{113}+2 K_{122}+K_{23}
\nonumber\\
&&-12 K_{113}-3 K_{115}-9 K_{124}-13 K_{223}-K_{25}-3 K_{34}
\Big]\,
 j_1(\kappa_1+\kappa_2+\kappa_3)
\nonumber
\\
&&
+\frac{1}{K_{333}}
\big\{
\big[K_2+K_{13}+K_4-\left(2 K_3+K_{12} \right) \kappa_3 
\nonumber
\\
&&
+
\left( K_2+K_{11}+K_4-K_{13}-4 K_{22}
\right) \kappa_3^2+\left( 2 K_3+2K_{12}-2 K_1+2\right)\kappa_3^3
\nonumber
\\
&&+
\left( 1+K_{11}+2K_2\right) \kappa_3^5-2 K_1 \kappa_3^5-3 \kappa_3^6
\big]
\frac{j_0(\kappa_1+\kappa_2-\kappa_3)}{\kappa_3} +{\rm perms}
\big\}
\nonumber
\\
&&+\big\{
\big[ K_{14}+3 K_{23}-K_{12}-K_3+\left( 2K_2-2 K_{22}-K_4-K_{13}\right) \kappa_3
\nonumber
\\
&&+\left(2 K_{12}-K_3+3 K_{23} +K_{14}\right)\kappa_3^2+
\left( 2 K_{22}+K_{13}-K_4-4 K_2-K_{11}
\right)\kappa_3^3
\nonumber
\\
&&-\left( K_{12}
+4 K_3+2 K_1\right)\kappa_3^4+\left( 9-K_{11}-4 K_2 \right) \kappa_3^5
+9\kappa_3^7
\big]
\frac{j_1(\kappa_1+\kappa_2-\kappa_3)}{\kappa_3} +{\rm perms}\big\}
 \nonumber
\\
{\cal C}_3&=&\frac{4\left(18 K_{122}+2 K_{113}-3 K_{12}-3 K_{3}\right)}{3\,K_{222}}
\,j_1(\kappa_1) j_1(\kappa_2) j_1(\kappa_3)\nonumber
\\
&&+\left[ \frac{8\left(
\kappa_3^3\,K_1+2 \kappa_3^2\,K_2-\kappa_3\,K_{12}
-3 \kappa_3^4
\right)}{K_{222}}
\,j_1(\kappa_1) j_1(\kappa_2) j_0(\kappa_3)+{\rm perms}\right]
\\
{\cal C}_4&=&\frac{16\,K_{11}}{K_{111}}\,j_1(\kappa_1)\,j_1(\kappa_2)
\,j_1(\kappa_3)
\eea
where $j_{0,1}$ are spherical
Bessel functions (see eq \eqref{defbes}).  The resulting bispectrum has a rich
shape and scale dependence, as discussed in the main text. 
%%%%%
\section{Computation of the  tensor-tensor-scalar bispectrum}
\label{appC}
%%%%%%

In this Appendix we evaluate  the tensor-tensor-scalar bispectrum
$B_{TTS}$ during radiation domination. Tensor modes are
 sourced at second order in perturbations by scalar fluctuations, which are enhanced
 at small scales by the non slowroll epoch described in sections \ref{sec_ps} and \ref{sec_bis}. This
 is a well studied phenomenon, and explicit semianalytic formulas for computing the properties of the induced tensor modes have been developed -- see e.g. \cite{Ananda:2006af,Baumann:2007zm,Kohri:2018awv}. We make use and develop
 previous results (especially \cite{Baumann:2007zm}) to compute the observable $B_{TTS}$
 during  radiation domination. For convenience, here we
 make use of dimensionful momenta $\vec k$ (their dimensionless
 version used in the main text can be  easily obtained as $\vec \kappa\,=\,-\tau_1 \,\vec k$).
 
 The Fourier transform of 
 tensor modes   sourced  by scalar fluctuations is 
\be
h_{\vec k}(\tau)\,=\,\frac{1}{a(\tau)} \int d \tilde \tau\,g_{k}(\tau, \tilde \tau)\,\left[ a(\tilde \tau)
S(\tilde \tau, \vec k)
\right]\,,
\ee
where $g_{k}(\tau, \tilde \tau)$ is a Green function.
In radiation domination it is
 given 
by 
\be
g_{k}(\tau, \tilde \tau)\,=\,\frac{1}{k}\left(\sin{(k \tau)}\sin{(k \tilde \tau)}
-\cos{(k \tau)}\cos{(k \tilde \tau)}
 \right)\,.
\ee 
The source contribution $S( \tau, \vec k)$ is given by a convolution integral involving
scalar fluctuations  \cite{Baumann:2007zm}
\be
\label{def_sou1}
S( \tau, \vec k)
\,=\,\int d^3 q \,{\bf e}(\vec k, \vec q)\,f(\tau, \vec k, \vec q)\,
\zeta_{\vec k-\vec q}\,\zeta_{\vec q}\,.
\ee
The curvature fluctuations appearing in eq \eqref{def_sou1}
are primordial perturbations evaluated right at the end of inflation, $\tau=0$. The 
quantity ${\bf e}(\vec k , \vec q )$ originates from contractions of polarization tensors.
It
is given by 
\bea
{\bf e}(\vec k, \vec q)\,=\, q^2-\frac{\left(\vec q \cdot \vec k\right)^2}{k^2}\,,
\eea
and it has the property ${\bf e}(\vec k ,\vec k- \vec q )\,=\,{\bf e}(\vec k , \vec q )$.
The quantity $f$ controls propagation effects from the end of inflation to time $\tau>0$
during radiation domination. In absence of anisotropic stress it is given by \cite{Ananda:2006af}
\be
f(\tau , \vec k , \vec q )\,=\,\,\left[ 
12 \,\Phi(\tau | \vec k-\vec q|) \Phi(\tau |\vec q|) 
+\left( 
\tau \, \Phi(\tau |\vec k-\vec q|)+\frac{\tau^2}{2} \Phi' (\tau |\vec k-\vec q|) 
\right)\Phi' (\tau |\vec q|) 
\right]\,.
\ee
In fact, the Bardeen scalar potential $\Phi_{\vec k}(\tau)$ 
during radiation domination is related to the curvature
perturbation at the end of inflation $\zeta_{\vec k}$ by the transfer function
\be
\Phi_{\vec k} (\tau)\,=\,\Phi\left(k \tau\right) \,\zeta_{\vec k}\,,
\ee
with 
\be
\Phi\left(k \tau\right)\,=\,\frac{2}{(k \tau)^2}\,\left(\frac{\sqrt{3}\,\sin{(k \tau/\sqrt{3})}}{ k \tau}
-\cos{(k \tau/\sqrt{3})} \right)\,.
\ee
The transfer function has the limit $\Phi\left(k \tau\right)\to2/3$ when
its argument tends to zero. 

\smallskip
By means of these quantities it is straightforward to compute 
$n$-point functions involving tensor  and/or scalar fluctuations.  
The two point function of tensor modes reads ($0\le \tau\le \tau_2$)
\be
\langle h_{\vec k_1}(\tau) h_{\vec k_2}(\tau)
\rangle\,=\,\frac{1}{a^2(\tau)}
\int_{\tau_0}^\tau\,d \tau_2\,\int_{\tau_0}^{\tau}
\,\d \tau_1\,a(\tau_1)\,a(\tau_2)\,
g_{\vec k_1}(\tau, \tau_1)\,
g_{\vec k_2}(\tau, \tau_2)\,
\langle
S( \tau_1, \vec k_1) S( \tau_2, \vec k_2) \rangle\,.
\label{twopote}
\ee
While the three point function with two tensors and one scalar mode
results
\bea
&&
\langle h_{\vec k_1}(\tau) h_{\vec k_2}(\tau)
\Phi_{\vec k_3}(\tau)
\rangle\,=\,
\nonumber
\\
&&
\frac{\Phi\left(k_3 \tau\right)}{a^2(\tau)}
\int_{\tau_0}^\tau\,d \tau_2\,\int_{\tau_0}^{\tau}
\,\d \tau_1\,a(\tau_1)\,a(\tau_2)\,
g_{\vec k_1}(\tau, \tau_1)\,
g_{\vec k_2}(\tau, \tau_2)\,
\langle
S( \tau_1, \vec k_1) S( \tau_2, \vec k_2) \zeta_{\vec k_3}\rangle\,.
\label{thrpote}
\eea
Using Wick's theorem, one finds
\bea
\langle S(\vec k_1, \tau_1) S(\vec k_2, \tau_2)
\rangle&=&
\delta(\vec k_1 +\vec k_2)
\nonumber
\\
&&
\times
\int d^3 q_1\,\left({\bf e}(\vec k_1, \vec q_1) \right)^2
\,f(\vec k_1, \vec q_1, \tau_1)\,  \left[ f(\vec k_1,\vec q_1, \tau_2)
+f(\vec k_1,\vec k_1- \vec q_1, \tau_2)
\right]
\nonumber
\\
&&\times\,{{ P}_\zeta( |\vec k_1-\vec q_1|)}%{|\vec k_1-\vec q_1|^3}
{{ P}_\zeta( q_1)}%{q_1^3}
\,,
\eea
where the curvature power spectrum is defined in eq \eqref{gedeP}. Plugging 
the previous expression in eq \eqref{twopote}, we can determine the
induced tensor spectrum.  It can be expressed as the convolution
\be
P_h(\tau, k)\,=\,\int d^3 q \,{{ P}_\zeta( |\vec k-\vec q|)}%{|\vec k_1-\vec q_1|^3}
{{ P}_\zeta( q)}\,{\cal F}_0(\tau,\vec k,\vec q)
\,,
\ee
with
\bea
{\cal F}_0(\tau,\vec k,\vec q)&=&
\frac{1}{a^2(\tau)}
\int_{\tau_0}^\tau\,d \tau_2\,\int_{\tau_0}^{\tau}
\,\d \tau_1\,a(\tau_1)\,a(\tau_2)\,
g_{ k}(\tau, \tau_1)\,
g_{ k}(\tau, \tau_2)\,
\nonumber
\\
&&\times
\,\left({\bf e}(\vec k, \vec q) \right)^2
\,f(\vec k, \vec q, \tau_1)\,  \left[ f(\vec k,\vec q, \tau_2)
+f(\vec k,\vec k- \vec q, \tau_2)
\right]\,.
\label{defFz}
\eea

The three point function in eq \eqref{thrpote} is more elaborated. 
An application of Wick theorem suggests to split the result in two separate convolution integrals
\be
\langle
S( \tau_1, \vec k_1) S( \tau_2, \vec k_2) \zeta_{\vec k_3}\rangle
\,=\,\delta\left( \vec k_1+\vec k_2+\vec k_3\right)\,\left(
I_1+I_2
\right)\,,
\ee 
with respectively
\bea
I_1&=&%\delta\left( \vec k_1+\vec k_2+\vec k_3\right)
\int d^3 q_1\,\Big[{\bf e}(\vec k_1, \vec q_1)
\,f(\vec k_1, \vec q_1, \tau_1)\,{\bf e}(\vec k_1+\vec k_3, - \vec q_1+\vec k_1 )
\,f(-\vec k_3-\vec k_1,  -\vec k_3-\vec q_1 , \tau_2)
\nonumber
\\
&&+{\bf e}(\vec k_1, \vec q_1)
\,f(\vec k_1, \vec q_1, \tau_1)\,{\bf e}(\vec k_1+\vec k_3,  \vec q_1+\vec k_3 )
\,f(\vec k_1+\vec k_3,  \vec k_1-\vec q_1 , \tau_2) \Big]
\nonumber
\\
&&\times{{\cal P}_\zeta( |\vec k_1-\vec q_1|)}%{|\vec k_1-\vec q_1|^3}
 \, B_\zeta\left( q_1,|\vec k_3+\vec q_1| ,  k_3
\right)\,,\\
I_2&=&%\delta\left( \vec k_1+\vec k_2+\vec k_3\right)\,
{\bf e}(\vec k_1,  \vec k_1+\vec k_3)\,f(\vec k_1,\vec k_1+\vec k_3, \tau_1)\,{{ P}_\zeta(  k_3)}%{|\vec k_3|^3}\,
 \nonumber
\\
&&\times
\int d^3 q_1\,{\bf e}( \vec k_1+\vec k_3, - \vec q_1)
\,f( \vec k_1+\vec k_3, -\vec q_1 , \tau_2)
 B_\zeta\left( |\vec k_1+\vec k_3|,|\vec q_1+ \vec k_1+\vec k_3|, q_1
\right)\,.
\eea
Accordingly, the $B_{TTS}$ bispectrum results (with $\vec k_2=-\vec k_1-\vec k_3$)
\bea
B_{TTS}(\vec k_1,\vec k_2,\vec k_3)&=&
\int d^3 q \,{{ P}_\zeta( |\vec k_1-\vec q|)}%{|\vec k_1-\vec q_1|^3}
\, B_\zeta\left( q,|\vec k_3+\vec q| ,  k_3
\right)\,{\cal F}_1(\tau,\vec k_1, \vec k_3, \vec q)
\nonumber
\\
&+&{{ P}_\zeta(  k_3)} \int d^3 q \,%{|\vec k_1-\vec q_1|^3}
 \,B_\zeta\left( |\vec k_1+\vec k_3|,|\vec q_1+ \vec k_1+\vec k_3|, q_1
\right)\,{\cal F}_2(\tau,\vec k_1, \vec k_3, \vec q)\,,
\label{totbisgw}
\eea
with 
\bea
%\\
{\cal F}_1(\tau,\vec k_1,\vec k_3, \vec q)&=&
\frac{1}{a^2(\tau)}
\int_{\tau_0}^\tau\,d \tau_2\,\int_{\tau_0}^{\tau}
\,\d \tau_1\,a(\tau_1)\,a(\tau_2)\,
g_{ k}(\tau, \tau_1)\,
g_{ k}(\tau, \tau_2)\,
\nonumber
\\
&&\times
\,\Big[{\bf e}(\vec k_1, \vec q)
\,f(\vec k_1, \vec q, \tau_1)\,{\bf e}(\vec k_1+\vec k_3, - \vec q+\vec k_1 )
\,f(-\vec k_3-\vec k_1,  -\vec k_3-\vec q , \tau_2)
\nonumber
\\
&&+{\bf e}(\vec k_1, \vec q)
\,f(\vec k_1, \vec q, \tau_1)\,{\bf e}(\vec k_1+\vec k_3,  \vec q+\vec k_3 )
\,f(\vec k_1+\vec k_3,  \vec k_1-\vec q , \tau_2) \Big]
\label{defbf1}
\\
{\cal F}_2(\tau,\vec k_1,\vec k_3, \vec q)&=&
\frac{1}{a^2(\tau)}
\int_{\tau_0}^\tau\,d \tau_2\,\int_{\tau_0}^{\tau}
\,\d \tau_1\,a(\tau_1)\,a(\tau_2)\,
g_{ k}(\tau, \tau_1)\,
g_{ k}(\tau, \tau_2)\,
\nonumber
\\
&&\times
\,\Big[{\bf e}(\vec k_1,  \vec k_1+\vec k_3)\,f(\vec k_1,\vec k_1+\vec k_3, \tau_1)\,
\,{\bf e}( \vec k_1+\vec k_3, - \vec q)
\,f( \vec k_1+\vec k_3, -\vec q , \tau_2)
\Big]\,.
\nonumber
\\
\label{defbf2}
\eea

A  simplification occurs in the squeezed limit, when
the momentum of the scalar perturbation $\vec k_3\to0$. The function ${\cal F}_2$ is suppressed by 
 a small factor $k_3^2$ with respect
to ${\cal F}_1$. The latter, moreover, becomes coincident to  ${\cal F}_0$.
Consequently, for $\vec k_3\to0$, we find the relation
\bea
\lim_{k_3\to0} B_{TTS}(k,k,k_3)
\,=\,P_{\zeta}(k_3)\,\int d^3 q \,{{ P}_\zeta( |\vec k-\vec q|)}%{|\vec k_1-\vec q_1|^3}
{{ P}_\zeta( q)}\,\,\left[1-n_\zeta(q)\right]\,{\cal F}(\tau,\vec k,\vec q)
\eea
 whose interesting
physical implications are considered in section \ref{sec_gw}.
% The function
%${\cal F}_0$ is introduced in eq \eqref{defFz}. 
We used the fact that the scalar
bispectrum satisfies Maldacena consistency relation in the squeezed limit --
section \ref{sec_bis}.
Besides
the squeezed limit above,
these formulas  are exact and can be used to
study other shapes of the tensor-tensor-scalar bispectrum during the
 radiation
dominated era.

\end{appendix}

{\small
%\addcontentsline{toc}{section}{References}
%\bibliographystyle{utphys}

%\bibliographystyle{utcaps}
%\bibliographystyle{kp}

%\bibliography{SYMMETRYrefs}
\providecommand{\href}[2]{#2}\begingroup\raggedright\endgroup

}

\end{document}